\documentclass[twocolumn,floatfix,prx,aps,eqsecnum]{revtex4-1}
\usepackage{float}
\usepackage{graphicx}
\usepackage{amsmath}
\usepackage{amssymb}

\usepackage[pdftex,bookmarks=false,colorlinks,linkcolor=blue,citecolor=blue,urlcolor=blue]{hyperref}

\usepackage{color}

\begin{document}
\title{Coupled Wire Model of $Z_4$ Orbifold Quantum Hall States}

\author{Charles L. Kane}
\affiliation{Department of Physics and Astronomy, University of
Pennsylvania, Philadelphia, PA 19104}
\author{Ady Stern}
\affiliation{Department of Condensed Matter Physics, The Weizmann
Institute of Science, Rehovot 76100, Israel}

\begin{abstract}

We introduce a coupled wire model for a sequence of non-Abelian quantum
Hall states that generalize the $Z_4$ parafermion Read Rezayi state.  The $Z_4$
orbifold quantum Hall states occur at filling factors $\nu = 2/(2m-p)$
for odd integers $m$ and $p$, and have
a topological order with a neutral sector characterized by the
orbifold conformal field theory with
central charge $c=1$ at radius $R=\sqrt{p/2}$. When
$p=1$ the state is Abelian. The state with $p=3$ is the $Z_4$ Read Rezayi state, and the series of $p\ge 3$
defines a sequence of non-Abelian states that resembles the Laughlin sequence.
Our model is based on clustering of electrons in
groups of four, and is formulated as a two fluid model in which each wire
exhibits two phases: a weak clustered phase, where
charge $e$ electrons coexist with charge $4e$ bosons and a strong clustered phase
where the electrons are strongly bound in groups of 4.   The transition between these two phases
on a wire is mapped to the critical point of the 4 state clock model, which in turn
is described by the orbifold conformal field theory.
For an array of wires coupled in the presence of a perpendicular magnetic field,
strongly clustered wires
form a charge $4e$ bosonic Laughlin state with a chiral charge mode at the edge, but no neutral mode and a gap for single electrons.
Coupled wires near the critical state form quantum Hall states with a gapless neutral mode described by the orbifold theory.
The coupled wire approach allows us
to employ the Abelian bosonization technique to fully analyze the physics of
single wire, and then to extract most topological properties of the
resulting non-Abelian quantum Hall states. These include the list of
quasiparticles, their fusion rules, the correspondence between bulk
quasiparticles and edge topological sectors, and most of the phases
associated with quasiparticles winding one another.

\end{abstract}

\maketitle

\section{Introduction}
\label{sec:I}

Recent works have studied the two dimensional quantum Hall effect as a
set of coupled planar parallel
quantum wires subject to a perpendicular magnetic
field\cite{kml2002,teokane2014,neupert2014,PhysRevB.90.115426,
sagi2014,santos2015,meng2015,sagi2015,huang2016,iadecola2016,PhysRevX.5.011011,PhysRevLett.116.036803,
PhysRevLett.116.176401,Fuji2017}. The easiest case to consider is that
of the
integer quantum Hall effect, where interactions between electrons are
not essential. The fractional
quantum Hall states require interactions, and the coupled
wire description enables the application of bosonization
techniques\cite{haldane81jpc,haldane81prl} for the analysis of these
interactions. As expected, among the fractional quantum Hall states the
Laughlin $\nu=1/m$ ``magic
fractions"\cite{Laughlin83} are easiest to handle, with the complexity
increasing when dealing with hierarchy states.
The non-Abelian quantum Hall, including the Moore Read
state\cite{mooreread1991} and Read Rezayi states
states\cite{readrezayi1999} were reproduced by coupled wire
constructions\cite{teokane2014},
but at the cost of introducing a spatially modulated
magnetic field.

In our earlier paper\cite{ksh2017}, to which the present paper is a
companion, we showed how to use a coupled wire
model to construct non-Abelian states that are a result of clustering of
electrons into pairs. These
states, of which the best known is the Moore-Read Pfaffian
state\cite{mooreread1991}, may also be described as various
types of  $p$-wave superconductors of Chern-Simons composite
fermions\cite{readgreen2000}. Our construction combined the
two ingredients common to all Read-Rezayi non-Abelian quantum Hall states:
the clustering of electrons (in this case into pairs) and the
construction of an edge
made of a chiral charge mode that is a Luttinger liquid and a chiral
neutral mode that is described by
a Conformal Field Theory (CFT) of a fractional central charge. It did not
 require a modulated magnetic field.

In this work we focus on another set of non-Abelian states, in which
electrons cluster to groups of
four, and the neutral edge mode is described in terms of an orbifold
theory\cite{byb, ginsparg1988}. The Read-Rezayi series of
non-Abelian states\cite{readrezayi1999} is based on the construction of
clusters of $k$-electrons at filling factors
$\nu = k/(mk+2)$ (with $m$ odd) or the clustering of $k$-bosons
at $\nu=k/(mk+2)$ (with $m$ even). In both cases it may be viewed as a
Bose condensate of these clusters, which, due to Chern-Simons flux
attachment, may be mapped onto
Bosons at zero magnetic field. The Read-Rezayi series span all positive
integer values of $k$.

\begin{figure}
\includegraphics[width=3in]{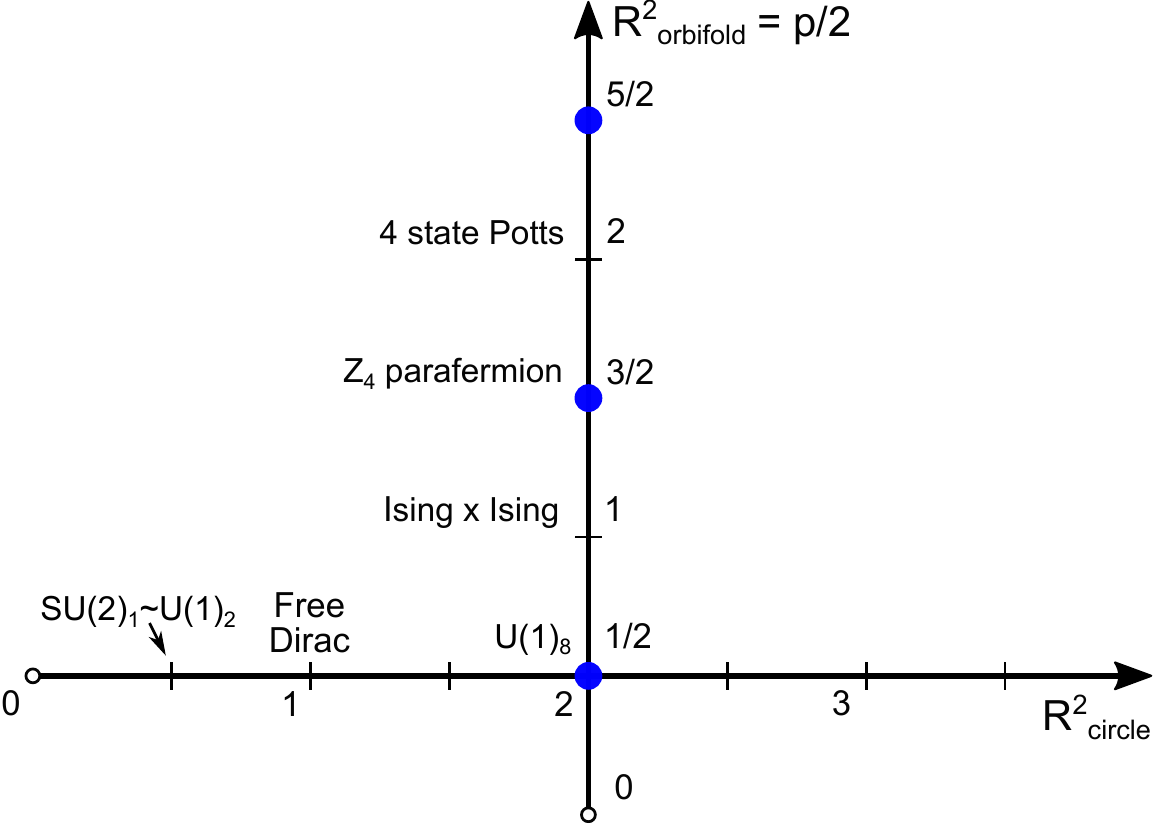}
\caption{Conformal field theories with $c=1$ include two intersecting lines of continuously varying critical points\cite{ginsparg1988b,ginsparg1988}.  The horizontal line describes a free boson compactified on a circle of radius $R_{\rm circle}$, while
the vertical line describes a free boson compactified on an orbifold of radius $R_{\rm orbifold}$\cite{Note1}.  The circle theory at $R_{\rm circle}=\sqrt{2}$ and the orbifold theory at $R_{\rm orbifold} = 1/\sqrt{2}$ are equivalent and related by an $SU(2)$ symmetry.
 The $Z_4$ orbifold states studied in this paper form a sequence analogous to the Laughlin sequence, and have edge states with a neutral sector described by the orbifold theory at $R_{\rm orbifold} = \sqrt{p/2}$, where $p$ is an odd integer, indicated by the solid blue circles.}
\label{c=1fig}
\end{figure}

The case of $k=4$ is unique. On one hand, it is too complicated to allow
for a quadratic mean field
Hamiltonian description. On the other hand, we show here that it does
allow for a rather detailed and
transparent analysis of its many-body Hamiltonian. Our work highlights the
connection between the
coupled wire model and the $c=1$ orbifold theory developed by Dijkgraaf,
Vafa, Verlinde and Verlinde\cite{dijkgraaf1989}, which formed the basis for the
analysis of
orbifold quantum Hall states carried out by Barkeshli and
Wen\cite{BarkeshliWen2011}.

The space of conformal field theories with  $c=1$ was studied extensively in the 1980's\cite{ginsparg1988b,ginsparg1988,dijkgraaf1989,byb}, and has the structure depicted in Fig. \ref{c=1fig}.     It includes two intersecting lines of continuously varying critical points, denoted the ``circle" line and the ``orbifold" line.   The circle line is equivalent to the theory of an ordinary single channel Luttinger liquid, which can be described as a free boson $\varphi$ with Lagrangian density $(\partial_\mu \varphi)^2/8\pi$ compactified on a circle, so that $\varphi \equiv \varphi + 2\pi R_{\rm circle}$\footnote{Here we use the convention of Ref. \onlinecite{ginsparg1988} for the normalization of $R_{\rm circle}$ and $R_{\rm orbifold}$\label{footnote1}}.  The radius $R_{\rm circle}$ is related to the Luttinger parameter $K$,
and specific radii describe rational CFT's of interest. The value  $R_{\rm circle} = 1/\sqrt{2}$ is the theory of the spin sector of $SU(2)$ fermions, described by $SU(2)_1$, or equivalently $U(1)_2$, with $K=2$.   The edge states of bosonic Laughlin states at filling $\nu = 1/m$ with $m$ even are described by $R_{\rm circle} = \sqrt{m}/2$.    Fermionic Laughlin states at $\nu = 1/m$ are described  by the circle theory at $R_{\rm circle} = \sqrt{m}$ with a constrained Hilbert space.   In particular, for $\nu=1$, the free Dirac fermion is at $R_{\rm circle} = 1$.

The orbifold theory is a variant on the Luttinger liquid model, and
describes a free boson compactified on a circle with radius $R_{\rm orbifold}$ in which angles $\theta$ and $-\theta$ are identified.   The orbifold theory at $R_{\rm orbifold}=1/\sqrt{2}$ and the circle theory at $R_{\rm circle}=\sqrt{2}$
(describing $U(1)_8$ - or $K=8$) are equivalent and are related by a hidden $SU(2)$ symmetry, which will play a key role in our analysis.   The orbifold theory at specific radii correspond to theories of interest, including doubled Ising model at $R_{\rm orbifold} = 1$, the $Z_4$ parafermion CFT at $R_{\rm orbifold}=\sqrt{3/2}$ and the 4 state Potts model at $R_{\rm orbifold}=\sqrt{2}$.

The $Z_4$ orbifold states that we will construct in this paper occur at filling factors
\begin{equation}
\nu = \frac{2}{2m-p}
\end{equation}
where $m$ and $p$ are odd integers.   They have neutral edge modes that are described by the orbifold CFT at  $R_{\rm orbifold} = \sqrt{p/2}$.    When $p=1$, the state is Abelian, and has an alternate description in terms of the circle CFT.   When $p=3$ the orbifold state is equivalent to the $Z_4$ parafermion Read Rezayi state.
Our coupled wire formulation takes advantage of the $SU(2)$ symmetry mentioned above, which allows for a description of the orbifold CFT in terms of Abelian bosonization.
This highlights the similarity between the sequence of orbifold states states for $p = 1, 3, 5, ...$ and the Laughlin sequence of Abelian states at $\nu = 1/m$, and allows a rather detailed analysis of the topological structure of the ground state and quasiparticle excitations.

The rest of the paper presents our analysis. Sec. (\ref{sec:II})
presents our results and the
physical picture that we develop to understand them. Sec.
(\ref{sec:III}) analyzes a single wire where the interaction between
electrons favors a clustering to $k$ electrons. Sec.  (\ref{sec:IV})
focuses on the case $k=4$ and shows how this case may be solved by
exploiting a hidden $SU(2)$ symmetry. Section (\ref{sec:V}) constructs
quantum Hall states from single wires of the type discussed in Sec.
(\ref{sec:IV}). Sec. (\ref{sec:VI}) analyzes the quasi-particles of
these states, and Sec. (\ref{sec:VII}) gives a concluding discussion.

\section{Physical picture and summary of results}
\label{sec:II}

\subsection{Single Wire}  \label{sec:II.A}
\subsubsection{General set-up} \label{sec:II.A.1}

Our approach for creating a coupled wire description of the
$k=4$ states is similar to our earlier construction of $k=2$
states\cite{ksh2017}. It is a two-fluid model, both of the single wire and
of the entire system.  For each wire we start with
two pairs of counter-propagating gapless modes: one carries clusters of
four electrons and is
described by the fields $\varphi_4,\theta_4$; the other   carries single
electrons and is described by
the fields $\phi_{1L},\phi_{1R}$. We then introduce two interaction
terms, one ($u$) that
back-scatters single electrons and one ($v$) that composes and
decomposes clusters into four
electrons. The Hamiltonian density for a single wire
then takes the
form
\begin{equation}
{\cal H}={\cal H}_0+{\cal H}_{int}
\label{HamGeneral}
\end{equation}
where ${\cal H}_0$ is the Luttinger liquid Hamiltonian density for the two
pairs of fields, and the interaction Hamiltonian density is
\begin{equation}
{\cal
H}_{int}=u\cos{(\varphi_4-2\phi_{1L}-2\phi_{1R})} +
v\cos{(\phi_{1R}-\phi_{1L})}\label{ham-original-dof}
\end{equation}
When the second interaction term dominates there is an energy gap for
single electron excitations, and the
system is in a strongly clustered state. It carries a pair of
counter-propagating gapless cluster
modes. When the first term dominates the single wire is in a
weakly-clustered
state in which one of the two pairs of counter-propagating modes is
gapped. In this state there is no
energy gap for single electron excitations. The operator that inserts a
single electron with a
vanishing energy cost does so with the insertion of a winding to the
phase of the bosonic phase field
of the clusters. Finally, in between these two phases there is a
critical state in which none of the modes is gapped\cite{lgn2002}.

\subsubsection{A  wire at the critical state}\label{sec:II.A.2}

The nature of the critical state is what makes the $k=4$ state unique
when compared to $k=2,3$. While
for $k=2,3$ there is a single critical point, for $k=4$ there is a
critical line, i.e., the low energy properties depend on the value of
$u$ and $v$. Furthermore, while
for $k=2,3$ the central charge of the gapless state is fractional, for
$k=4$ it is an integer.

For the $k=4$ case the competition of the $u,v$ terms in
(\ref{ham-original-dof}) is reminiscent of the quantum four-state
clock-model\cite{Fradkin1980}, composed of one dimensional lattice of
``clocks". This model is a special case of the more general Ashkin-Teller
model\cite{ginsparg1988,byb}, which has recently appeared in a number of contexts \cite{lgn2002,ZhangKane2014,KaneZhang2015,meidan2017}.   In this
model each site hosts a phase degree of freedom $\phi_i$. An interaction
term assigns an energy cost
$\sim\cos{4(\phi_i-\phi_{i+1})}$ to a phase difference between
neighboring site, while an on-site term
introduces a change of the local phase $\phi_i\rightarrow
\phi_i\pm\frac{\pi}{2}$. The relative size
of these terms determines whether the system is in an ordered or
disordered phase. As we describe
below, the critical state of the single wire of our problem has
much similarity with the critical
state of the Ashkin-Teller model. Like the latter,  its low energy
spectrum includes an orbifold theory of central charge $c=1$, whose
properties are analyzed in detail in Ref. \onlinecite{dijkgraaf1989}.

Our focus here is on  wires in the critical state. For
all values of $k$ the kinetic energy of the two counter-propagating
pairs of modes is quadratic in the
bosonic fields that describe the modes, while the two competing
interaction terms are cosines of
combinations of the bosonic fields, which do not commute with one
another. The non-commutativity of
the interaction terms makes the Hamiltonian generally difficult to
handle. For $k=2$ the fermionic
language comes to the rescue, since the interaction turns out to be
quadratic in terms of properly
chosen Majorana fermions. For $k=4$ the interaction terms are not
quadratic. However, the $k=4$ case has a hidden $SU(2)$ symmetry, which is
not apparent in Eq. (\ref{ham-original-dof}). Due to that symmetry, a
properly
chosen fermionic representation allows for an expression of the
interactions as interactions of small momentum transfer,
which allows for their mapping onto a quadratic Luttinger liquid. The
nature of the mapping imposes constraints
on the Luttinger liquid, and these constraints translate into an
orbifold theory.

Our study hops between the fermionic and bosonic representation of the
one dimensional degrees of
freedom that we analyze.
The bosonization approach to one dimensional systems allows for the
definition of vertex operators of
the form $e^{i{\bf \alpha\cdot \phi}}$ where ${\bf\phi}\equiv\left
(\phi_{e,R},\phi_{e,L},\phi_4,\theta_4\right )^T$ is the vector of
bosonic fields that describe the
system and $\bf\alpha$ is a vector of real numbers. The value of
$\bf\alpha$ determines the quantum
statistics of the operator. When the components of $\alpha$ are all
integers, the operator is local,
i.e., it is composed of creation and annihilation operators of single
electrons and four-electron
clusters within a localized region. All operators within a Hamiltonian
must obviously be local. In
the case we consider, where the starting point is that of two pairs of
counter-propagating bosons, the hidden $SU(2)$ symmetry is brought to the forefront
by choosing a set of four vectors $\bf \alpha$ that express the problem
in terms of two
types of fermions, each having left and right moving
branches. When we assign the two
types of fermions a fictitious spin ``up"/``down", any operator that
involves the two types of fermions
takes the form of a spin-$1/2$ field. From here on we will use the term
spin freely, referring always
to the fictitious spin. We will refer to these newly defined fermions as
``the $SU(2)$ fermions", to
distinguish them from the original  (spinless) electronic degrees of freedom.
The one dimensional $SU(2)$
fermions may be described by two pairs of counter propagating bosonic
modes, which we denote by
$\phi_{s,a}$, with $s=\uparrow,\downarrow$ denoting the spin
direction and $a=L,R$ denoting the
direction of motion.

The expression of the original degrees of freedom in terms of $SU(2)$
fermions is possible for all values of
$k$. It is useful for $k=4$ due to three unique
characteristics\cite{lgn2002, ginsparg1988}. First, there is a simple
criterion
that determines whether an operator expressed in terms of the $SU(2)$
fermions is local in terms of the
original electrons. This criterion states that a local operator is an
operator that changes the
number of spin-down fermions by an even number. Thus, the physical
subspace of the Hilbert space of
the $SU(2)$ fermions is constrained to the states at which the number of
down fermions is even, and the
local operators commute with the parity of the number of spin-down
fermions.  Second, many of the
operators that are local in terms of the original electrons turn out to
be local also in terms of the
$SU(2)$ fermions (exceptions will be elaborated on below). And third, at
the critical point the
Hamiltonian takes a particularly simple form in terms of the $SU(2)$
fermions. The kinetic term is an
isotropic ferromagnetic coupling which does not mix different
chiralities. It is quadratic in the
$\phi_{s,a}$'s. The critical interaction term couples the $x$-components
of the spins of right and left
moving $SU(2)$ fermions to one another. So, not only is the Hamiltonian
local with respect to the
fermions, it is also in a form that may be diagonalized. Its
diagonalization, however, requires us to
bosonize the $SU(2)$ fermions, since the Hamiltonian is quartic in $SU(2)$
fermion operators.

In the bosonized language of the $SU(2)$ fermions, the $x,y$-components of
the spin density are
non-linear in the bosonic fields $\phi_{s,a}$, involving factors such
as $S^\pm \sim \exp{\pm
i(\phi_{a,\uparrow}-\phi_{a,\downarrow})}$. In contrast, the
$z$-component is linear, involving
only $S_z \sim \partial_x(\varphi_{a,\uparrow}-\varphi_{a,\downarrow})$.
Consequently, it is desirable to
rotate the spin axes by $\pi/2$ around the $y$-axis, such that
the coupling of $x$-components
of the right- and left-moving spins becomes a coupling of the
$z$-components. Were it not for the
constraint imposed on the physical subspace, this would have been just a
renaming of axes. However,
the rotation affects also the constraint, transforming it to the
statement that in the rotated frame a
local operator is an operator that is invariant to the interchange of
spin-up with spin-down fermions.

When the transformation from the original degrees of freedom to the
rotated fermions is completed, the
effect of the critical interaction  is to transform the two
pairs of counter-propagating
bosonic modes, of the clusters and the single electrons, into two pairs
of coupled counter-propagating
modes, of the spin-up and down $SU(2)$ fermions, subjected to a constraint
on the allowed operators and
allowed states. The gapless modes of the rotated fermions can then be
described by a third and last
set of bosonic fields $\chi^r_{\rho,a}$ and $\chi^r_{\sigma,a}$, where
the super-script $r$ indicates
the rotated frame, the subscripts $\rho,\sigma$ indicate charge and spin
fields, and the subscript $a$
indicates again a direction of motion. The Hamiltonian is quadratic in
these fields, and the
interaction term couples only the spin fields. The Hamiltonian is
diagonalized to a pair of
counter-propagating charge modes and a pair of counter-propagating spin
modes, with the only parameter
in the diagonalization being the relative strength of the critical
interaction to the kinetic term.
This parameter determines the relative velocity of the charge and spin
modes, as well as the
eigen-operators of the spin mode. The eigenmodes mix the right- and
left-moving fermions of the
non-interacting problem to create chiral eigenmodes of the interacting
one. There is a set of discrete
values of the critical interaction parameter $\lambda$ for which the
eigen-operators of the spin modes create  an integer number $p$ fermions
of one
chirality and $p-1$ fermions of the opposite chirality.  This discrete
set of $\lambda$'s play a
special role below. The most obvious example is the non-interacting
case, $\lambda=0$, for which $p=1$. Operators are local when they are
invariant to the transformation
$\chi^r_{\sigma,a}\rightarrow -\chi^r_{\sigma,a}$ for both $a=R,L$. The
charge mode is not affected by
this constraint.

As mentioned before, the transformation from the four electron clusters
and single electrons to the
$SU(2)$ fermions has the virtue that almost all local operators in the
original degrees of freedom
correspond to local degrees of freedom in the fermionic representation.
Notable exceptions are the operators
$e^{i(2n+1)\theta_4}$, with an integer $n$. In the original degrees of
freedom, the $n=0$ operator is an operator
that introduces a $2\pi$ kink into the field $\phi_4$. When expressed in
terms of the $SU(2)$ fermions,
it becomes an operator that introduces a $\pi$ kink into the spin field
of the fermions.

\subsection{From coupled wires to a quantum Hall state}\label{sec:II.B}

 Tunneling between wires forms a quantum Hall state when it gaps the
 gapless modes in the bulk and
 leaves gapless chiral modes near the edge\cite{kml2002,teokane2014}. In
 an idealized situation, that would happen when the
 tunneling operator couples only left movers of one wire to right movers
 of a neighboring wire. In the
 present case each wire has two counter-propagating pairs of modes. In
 the strong-clustered  and
 weak-clustered phases one of these pairs is gapped by intra-wire
 interactions, such that inter-wire
 tunneling needs to gap only one pair. In the critical state, however,
 two tunneling terms are needed
 to gap the two pairs of gapless modes.

 To be effective, the tunneling terms should satisfy several conditions:
 there must be a spectral
 weight for the tunneling particle to tunnel into or out of a wire at
 the chemical potential; the
 tunneling particle must be local; and there must be a momentum balance.
 The sum of the momentum that the
 tunneling particle takes from its wire of origin and the momentum that
 it receives from the Lorenz
 force when it tunnels should equal the momentum that is associated with
 the state to which it tunnels
 in the wire of destination.  These requirements are general to all wire
 constructions of quantum Hall
 states, but some aspects of their application are  unique to the
 present context.

 The identity of the particles that have a spectral weight to tunnel  at
 the chemical potential
 depends on the phase that the single wire is in. In the strongly
 clustered states, the only particles that may tunnel are clusters of
 four electrons, which in the $SU(2)$ language are described as two
 spin-up and two
 spin-down fermions. Thus, each cluster is spinless.
 In the critical
 state all particles can tunnel at the chemical potential.

 The notion of locality appears here twice. The tunneling particle  must
 be local in terms of the
 electrons and the 4-electron clusters. However, it does not  have to be
 local in terms of the $SU(2)$
 fermions, which are calculational constructs. It is to be expected,
 though, that tunneling terms that
 are local also in terms of the $SU(2)$ fermions would be easier to
 analyze. Indeed, we study quantum
 Hall states based on such terms here, and defer those for which the
 tunneling terms are non-local in
 terms of the $SU(2)$ fermions to a future publication.

 The balance of momentum is the major factor that determines the filling
 factors for which quantum
 Hall states are formed. The filling factors formed are those for which
 when a charge $q$ tunnels
 between states at the chemical potential on different wires, the
 momentum it adds to the electronic
 system in the wires equals the momentum it receives from the Lorenz
 force, namely $qBd$, where $d$ is
 the inter-wire distance. For $\nu=1$, for example, a single electron
 tunnels between two Fermi
 points, such that $eBd=2k_F=2\pi n_e$, which corresponds to $\nu=1$
 (here $n_e$ is the
 one-dimensional electron density). When the tunneling event is
 accompanied by two intra-wire $2k_F$
 scattering events, one in each of the participating wires, the momentum
 balance is $eBd=6k_F$, and
 the $\nu=1/3$ state is obtained.

For the present problem the condition for momentum balance depends on
the identity of the tunneling
particle, which depends on the state of the individual wires.
Tunneling operators of 4-clusters
have the form
\begin{equation}
e^{i\left(\varphi_4 + L \theta_4 + M (\phi_{1R}-\phi_{1L})\right)}
\label{4tunneling}
\end{equation}
Here and below $L,M$ are integers, and $n_b$ is the number density of
the bosonic clusters. These
operators involve a momentum of $2\pi (Ln_b+Mn_e)$. When the wires
are in the strongly clustered
phase, the only active degrees of freedom are the bosonic $4e$ clusters,
such that $M=0$. Based on
tunneling operators with $L=l$  Laughlin states of cluster filling
factor $\nu_{4e}=1/{2l}$ may be
formed, which correspond to electronic filling factors of $\nu=8/l$. The
states are Abelian, and the
edge carries a single chiral mode, with a central charge of one.

When the wires are in their critical states there are two
tunneling processes, aimed at gapping the charge and the neutral modes.
The charge mode is gapped by the same operator that gaps it in the
strongly
clustered state -  a tunneling of a four-electron cluster, which
corresponds to the tunneling of two
spin-up and two spin-down $SU(2)$ fermions. The charge mode is insensitive
to the interaction strength
$\lambda$, and hence so is also the operator that gaps it.

The operator that gaps the spin mode  must tunnel charge, so
that it may get momentum from the Lorenz force, and must carry spin, to
couple to the spin mode. A
single electron tunneling is then the natural candidate.
Generally, the single electron tunneling operator is
\begin{equation}
e^{i\left( \frac{1+m}{2}\phi_{1R} + \frac{1-m}{2}\phi_{1L} +  l \theta_k
\right)} .
\label{1tunneling}
\end{equation}
with $m$ being an odd integer, and $l$ an integer.
The momentum involved is $2\pi (mn_e+ln_b)$. Close to the transition
$n_b\gg n_e$, such that $4en_b$ is
approximately the total charge density,  and the momentum balance
condition  $2\pi ln_b=eBd$
translates to $\nu=2/l$. Furthermore, when expressed in terms of the
$SU(2)$ fermions, the operators
(\ref{1tunneling}) are local only for odd $l$, imposing the final
restriction on our analysis to
filling factors of the form $\nu=2/l$, with odd $l$.

The momentum balance, and hence the filling factor, do not depend on the
value of $m$ in
(\ref{1tunneling}). This value is determined by the requirement that the
single electron tunneling
term gaps the spin mode. Since the eigenvectors of the spin mode couple
right- and left-moving
electrons in a way that depends on the interaction scale $\lambda$, the
tunneling operator should
depend on the interaction as well. As explained before, for a set of
discrete value $\lambda(p)$ there
is a single electron operator that couples only to one chirality of the
spin mode. For odd $p$, this
happens when $m=(l+p)/2$.

Expressing $l$ in terms of $m$ and $p$, we summarize this subsection by
saying that the analysis we
present in this paper takes us from the Abelian quantum Hall states at
$\nu=2/(2m-p)$, which are
formed by wires at their strongly clustered phase, to quantum Hall
states at the same $\nu$ formed by wires at their critical phase.
The phase boundary between the two states is determined by
the ratio of intra-wire single electron back scattering and inter-wire
single electron tunneling. As
we explain in the next subsection, the latter states are non-Abelian.

As a final remark on the subject we note that when the wires are in
their weakly clustered phase,
single electron tunneling may lead to the formation of an anisotropic
quantum Hall state. A chiral
charge mode then runs along the entire edge, and a neutral achiral edge
mode exists along edges that are not
parallel to the wires. The properties of this state are quite similar to
those of its $k=2$
counterpart\cite{ksh2017}.

\subsection{Topological properties of the quantum Hall
states}\label{sec:II.C}

The fractionalized quasiparticles of fractional quantum Hall states are
manifested in the ground
state degeneracy in a torus geometry, in the different topological
sectors of the chiral edge modes in
an annular geometry,  and as gapped bulk excitations. Within the wire
construction, the first two
manifestations are expressed in terms of operators that create
quasiparticle-quasihole pairs and
position them on the two edges of the annulus or take them around the
torus. The bulk
quasiparticle-quasihole pair occurs as a kink-antikink pair in a bosonic
phase variable that is pinned
to one of several degenerate values in the system's
bulk\cite{kml2002,teokane2014}. In Sec. (\ref{sec:VI}) we analyze both
edge and
bulk quasiparticles. Here we describe the physical picture of bulk
quasi-particles. We do so using the
language of the $SU(2)$ fermions in the rotated system of axes.

The quasiparticle content of the strongly clustered quantum Hall states
is rather easy to understand.
The states we consider here are Laughlin states of 4-electrons clusters
with the cluster filling factors being $1/8l$,
and $l$ being odd integer. Their $K$-matrix is the number $8l$, and their
charge vector is the number $4$.
As such, they carry $8l$ fractionally-charged quasi-particles of charges
$Q/2l$, with $Q=1,...,8l$.
Within the wire construction, single electron backscattering in
each wire gaps the spin degree of
freedom and makes it irrelevant to the quantum Hall physics, while
cluster tunneling between
neighboring wires gaps the charge degrees of freedom, except one chiral
mode at each edge. Within a
bosonized description of the $SU(2)$ fermions, cluster tunneling leads to
the pinning of a particular
relative phase of the charge modes of neighboring wires, which we denote
by ${\bar \theta}_{\rho}$ and
define precisely in Eq. (\ref{barthetabarphirho}). This relative phase
is pinned to one of $8l$ possible values for
which the energy is minimal and degenerate. The quasi-particles reside
between wires, in the form of
kinks in the pinned relative phase. The charge $Q$ of the quasi-particle
is coupled then to the phase
jump between the start point and the end point of the kink, which is
${\bar \theta}_{\rho}=Q\pi/4l$.

When the quantum Hall state is formed of wires at the critical
state
the spin modes are gapped by single electron tunneling terms between
neighboring wires. These terms
involve both the charge and the spin modes, and couple them in an
interesting way: when
${\bar\theta}_\rho$ is pinned to $j\pi/4l$ with an even value of $j$, it
pins a relative phase of the
spin sector ${\bar\theta}_\sigma$ (defined precisely in Eq.
(\ref{barthetaphivarsigma})) to one of $p$ possible values
which are evenly spaced.
When the value of $j$ is odd, a different relative phase
${\bar\varphi}_\sigma$ (defined precisely in Eq.
(\ref{barthetaphivarsigma})) is pinned to one of $p$ values which are
evenly spaced.
The two phases ${\bar\varphi}_\sigma,{\bar\theta}_\sigma$ do not commute
with one another. As a
consequence, kinks in ${\bar\theta}_\rho$ of even $j$ may come together
with kinks in the pinned spin
phase, be it ${\bar\varphi}_\sigma$ or ${\bar\theta}_\sigma$. In
contrast, kinks of odd $j$ come together
with an excitation similar to the one occurring when two
counter propagating FQHE edge modes
are gapped alternately by a superconductor and by
backscattering\cite{Lindner2012, Clarke2013, Vaezi2014, Meng2012,
Mong2014}. The appearance of this excitation makes the quasiparticle
associated with an odd value of $j$ non-Abelian for all values of $p>1$.
Excitations of this type will be referred to as a twist fields, and
will be denoted by $\sigma,\tau$.

The interface between two regions with pinned non-commuting phase
variables is one of two sources for
non-Abelian quasiparticles. The other source is the constraint imposed
on the Hilbert space of the
$SU(2)$ fermions. In the rotated basis, physical states should be
invariant to the interchange of
spin-up and spin-down fermions. This interchange may be expressed as a
transformation on the values of
the bosonic phases ${\bar\varphi}_\sigma,{\bar\theta}_\sigma$. The values
to which each of these phases
may be pinned are either invariant under the transformation or form
pairs that are transformed onto
one another. In the former case the states are allowed. In the latter
case, which happens only for $p\ge 2$, they may occur only as
superpositions of states pinned to the members of the pair. In these
cases  a phase variable is
in a superposition of two distinct values over
distances that may be macroscopic, the distances between a kink and its
inverse. The superposition is protected from decoherence by the
constraint, which makes operators that may distinguish the two
superposed values unphysical. Again, a
quasiparticle is a kink that separates between two regions with
different pinning values. When two
quasiparticles that create superpositions are fused, the constraint
forces the fusion to have several
possible outcomes, making the quasiparticles non-Abelian. These
quasiparticles will be referred to as $\Phi_\lambda$.

The imposition of the constraint on the Hilbert space of the spin mode
has another consequence - it
splits the vacuum sector of the spin mode into two topologically
distinct sectors, a topologically trivial vacuum and a non-trivial
neutral particle. A particle-antiparticle pair of the latter is created
by an operator that is invariant to the interchange of spin-up and
spin-down $SU(2)$ fermions of both chiralities, but is odd under this
interchange when carried out for one chirality only.

Our analysis of the constrained system of the $SU(2)$ fermions and its
coupling between different wires
reproduces for the non-Abelian states the entire set of primary fields
that is familiar from the
orbifold description of these states, and provides a bosonized
description for each of these operators\cite{dijkgraaf1989,byb}.

\section{Clustering Transition on a single wire}\label{sec:III}

In this section we consider in detail a single one dimensional wire with an
attractive interaction that favors the formation of $k$ particle bound
states.  Our approach is to develop a ``two fluid" model, described by a
two channel Luttinger liquid theory, that describes charge $ke$
particles coexisting with charge $e$ particles (which can be either
fermions or bosons).   We will show that for this wire there are two
distinct phases.  There is a ``strong clustered" phase, in which there
is a gap for the addition of a charge $e$ particle, so that single
particle Green's function decays exponentially.  In addition there is a
``weak clustered" phase in which the single particle gap vanishes, and
the Green's function has a power law decay.    These phases will be
identified as the ordered and disordered phases of a $Z_k$ clock type
model.   For $k<4$, there is a transition in the $k$ state Potts model
universality class.  For $k=4$, there is a line of critical points
characteristic of the Ashkin Teller model that maps to the orbifold
conformal field theory.   For $k>4$ there is an intermediate gapless
phase.

\subsection{Bosonization }\label{sec:III.A}

We begin by developing a model for clustering on a single one
dimensional wire.    Our strategy mirrors the approach of Ref.
(\onlinecite{ksh2017}), where pairing was implemented by coupling a charge
$e$ Fermi gas to a one-dimensional Luttinger liquid of charge $2e$
bosons.   Here we generalize this to allow for clusters of $k$
particles.    Our primary interest in this paper will be $k=4$, and for
simplicity we will assume here that $k$ is even, so that the clusters
are bosons.  However,
the model which we derive can also be applied for odd $k$, where the
clusters are fermions, as well as to the case where the charge $e$
particles are bosons.

We begin with a Hamiltonian density of the form
\begin{equation}
{\cal H} = {\cal H}^0_e+ {\cal H}^0_{ke} + {\cal H}^{\rm int}_c
\end{equation}
where
\begin{equation}
{\cal H}^0_e =   \psi_1^\dagger (\epsilon_0 - \partial_x^2/2m - \mu)
\psi_1
\label{h0e}
\end{equation}
describes a one dimensional system of non-interacting charge $e$
fermions, and
\begin{equation}
{\cal H}^0_{ke} = \frac{v}{4\pi}[g(\partial_x\varphi_{k})^2 +
\frac{1}{g}(\partial_x\theta_{k})^2 ] -  k \mu
(\frac{\partial_x\theta_{k}}{2\pi}+\bar\rho_{k})
\label{h0ke}
\end{equation}
describes a one-dimensional Luttinger liquid of charge $ke$ bosons with
average density $\bar\rho_{k}$.    The Luttinger liquid is characterized
by a Luttinger parameter $g$, and is expressed in terms of variables
that satisfy
\begin{equation}
[\theta_{k}(x),\varphi_{k}(x')] = 2\pi i \Theta(x-x').
\label{thetaphicomm}
\end{equation}
A charge $ke$ boson is created by $e^{i\varphi_{k}}$, and the number
density of charge $ke$ bosons is $\bar\rho_{k} +
\partial_x\theta_{k}/2\pi$.    We assume the charge $e$ and $ke$ sectors
are in equilibrium at a chemical potential $\mu$ and are coupled by a
clustering term of the form
\begin{equation}
{\cal H}^{\rm int}_c =   \Delta e^{i \varphi_{k}}  \prod_{j = 1}^k
(\partial_x^{j-1} \psi_1).
\label{hc}
\end{equation}
This term describes a local clustering interaction which turns $k$
charge $e$ fermions into a charge $ke$ boson and vice versa.  The
derivatives are necessary due to the Fermi statistics of $\psi$.  This
term is the generalization of a spinless p-wave BCS pairing term, when
$k=2$.   In addition, we will consider below additional forward
scattering interactions between the charge $e$ and charge $ke$ sectors.

When $\epsilon_0-\mu$ is large and positive, the charge $e$ particles
will be depleted, and all of the charge density will reside in the
charge $ke$ sector.   This strongly clustered phase is a gapless
Luttinger liquid of charge $ke$ particles that has a gap for adding
charge $e$ particles.     For $\epsilon_0-\mu$ large and negative, the
charge $e$ and charge $ke$ particles coexist.   For $k=2$ we showed in
Ref. \onlinecite{ksh2017} that this is a weakly paired phase, in which
there is no gap for adding charge $e$ particles, and we showed that the
transition between weak and strong pairing phases is in the 2D Ising
universality class.   We anticipate a similar structure here, but unlike
the $k=2$ case, there is no free fermion limit in which the problem is
solvable.   Here we will develop a different approach by bosonizing the
charge $e$ sector.

A difficulty with directly bosonizing $\psi$ in Eq. \ref{h0e} is that
the clustering transition occurs when the fermions are depleted.  It is
difficult to bosonize near the bottom of a band.   An alternative is to
consider a theory of Dirac fermions, or equivalently a finite density of
fermions in the presence of a commensurate periodic potential.   This
opens a gap at the Fermi energy, which has the same effect as depleting
the Fermion density.

We therefore replace Eq. \ref{h0e} by
\begin{equation}
\tilde{\cal H}^0_e =
-iv_F (\psi_{ 1R}^\dagger \partial_x \psi_{ 1R} - \psi_{1 L}^\dagger
\partial_x\psi_{ 1L})
+ \Gamma (\psi_{1R}^\dagger \psi_{ 1L} + h.c.)
\end{equation}
Here $\psi_{1R}$ and $\psi_{1L}$ are right and left moving chiral Dirac
fermion operators, subject to a backscattering term $\Gamma$, which
opens an insulating energy gap.
We replace the clustering interaction by
\begin{equation}
\tilde{\cal H}^{\rm int}_c =     \Delta e^{i\varphi_{k}
}\prod_{j=1}^{k/2} \left( \partial_x^{j-1} \psi_{1R} \right)\left(
\partial_x^{j-1} \psi_{1L} \right) + h.c.
\end{equation}

For $k=2$, this theory has the structure of a four band Bogoliubov de
Gennes theory for the transition between a trivial and topological
superconductor.    When $\Gamma \gg \Delta$, the fermions acquire a band
gap at the Fermi energy, and they are effectively depleted.   On the
other hand, when $\Gamma\ll\Delta$, the fermions are not gapped.   These
phases and the transition between them are the same as simpler two band
Read Green model described by (\ref{h0e},\ref{hc}) when $k=2$.     It is
natural to expect that the equivalence between these two models also
holds for more general values of $k$.

Importantly, in this alternative model we tune through the clustering
transition by varying the relative magnitudes of $\Delta$ and $\Gamma$,
not by varying the chemical potential.  We fix the chemical potential to
be precisely at the Dirac point, where the left and right moving
fermions have momentum zero.    Therefore, the total average density
$\rho_e = \rho_{1} + k \rho_k$ is entirely in the charge $ke$ sector, so
that $\rho_k = \rho_e/k$.

We now bosonize the zero momentum Dirac fermions by writing
\begin{equation}
\psi_{1 A}(x) \sim e^{i\phi_{1 A}(x)},
\end{equation}
where $A = R,L$ and the boson fields commute with $\varphi_k$,
$\theta_k$ satisfying
\begin{equation}
[\phi_{1A}(x),\phi_{1A'}(x') ] = i\pi {\rm sgn}(x_A-x'_{A'}) .
\label{phi1com}
\end{equation}
Here we have chosen a convention in which on a finite wire of length $L$
with periodic boundary conditions we specify an ordering for the fields
$\phi_{1A}(x)$.    We define $x_L = L-x$ and $x_R = L+x$.   Since $x_R >
x_L$, it follows that $[\phi_{1R}(x),\phi_{1L}(x')] = i\pi$.   This
ensures the proper anticommutation between $\psi_{1R}$ and $\psi_{1L}$.

Our theory is  characterized by four bosonic fields, which are
convenient to combine into a column vector
\begin{equation}
\Phi= (\phi_{1R}, \phi_{1L}, \theta_k, \phi_k)^T.
\label{phivec}
\end{equation}
The fields $\Phi_I$ are each defined modulo $2\pi$.  It follows that the local
operators in the theory can be expressed as derivatives of $\Phi_I$ and
as vertex operators of the form
\begin{equation}
e^{i{\bf N} \cdot \Phi}
\label{localop}
\end{equation}
where ${\bf N}$ is a four component integer valued vector.   The basic
vertex operators in the theory include the charge $e$ particle creation
operators $e^{i\phi_{1R/L}}$ and the charge $ke$ particle creation
operator $e^{i\varphi_4}$.   In addition, the operator $e^{i\theta_4}$
describes the phase of the modulation of the charge $ke$ density at wave
vector $2\pi \bar \rho_k$, analogous to the $2k_F= 2\pi\bar\rho_1$
density modulation $e^{i(\phi_{1R}-\phi_{1L})}$.  Equivalently,
$e^{i\theta_k}$ describes the tunneling of a flux $h/ke$ vortex in the
charge $ke$ boson order parameter across the wire.

The original electron  operator will in general be a sum of all charge
$e$ operators, which include the bare fermion operators $\psi_{1R/L}$
along with composite operators that include scattering from the density
fluctuations of the charge $e$ ($ke$) particles.
This has the form
\begin{equation}
\psi_1(x)  =  \sum_{ml}  c_{ml} \psi_{ml}  e^{i 2\pi\frac{l}{k}
\bar\rho_e  x}
\label{Psi1}
\end{equation}
where $c_{ml}$  are non universal constants and
\begin{equation}
\psi_{ml} = e^{i\left( \frac{1+m}{2}\phi_{1R} + \frac{1-m}{2}\phi_{1L} +
l \theta_k \right)} .
\label{psiml}
\end{equation}
Here $l$ is any integer and $m$ is an odd integer.
 In (\ref{Psi1}) we have used the fact that $\bar\rho_1=0$ and
 $\bar\rho_k = \bar\rho_e/k$.   The operators $\psi_{1 0}$ and
 $\psi_{-10}$  are the bare fermion operators $\psi_{1R/L}$, while the
 rest of the operators are composites that involve additional
 backscattering of the charge $e$ and/or $ke$ particles.    In general,
 $\psi_{-m-l}$ and $\psi_{ml}$ are related by a reflection that
 interchanges right and left movers.

Similarly, we can consider composite charge $ke$  operators of the form
\begin{equation}
\Psi_k(x)  =  \sum_{NL} C_{ML} \Psi_{ML} e^{2\pi i \frac{M}{k}
\bar\rho_e x}
\label{Psik}
\end{equation}
with
\begin{equation}
\Psi_{ML} =e^{i\left(\varphi_k + L \theta_k + M
(\phi_{1R}-\phi_{1L})\right)},
\label{PSIML}
\end{equation}
where $M$ and $L$ are integers.

Expressed in these variables, our Hamiltonian takes the general form
\begin{equation}
{\cal H} = {\cal H}_0 + {\cal H}_{uv},
\label{h1}
\end{equation}
with
\begin{equation}
{\cal H}_0 = \sum_{IJ}V_{IJ} \partial_x\Phi_I \partial_x\Phi_J
\label{h00}
\end{equation}
\begin{equation}
{\cal H}_{uv} =
u \cos(\frac{k}{2}(\phi_{1R}+\phi_{1L}) - \varphi_k) +
v\cos( \phi_{1R}-\phi_{1L} )
\label{hv}
\end{equation}
where $u$ and $v$ are proportional to $\Gamma$ and $\Delta$.   In ${\cal
H}_0$ we have included a general positive definite forward scattering
interaction matrix, which in addition to accounting for the Luttinger
parameter $g$ in (\ref{h0ke}) includes forward scattering interactions
that act in the charge $e$ sector as  well as coupling terms between the
charge $e$ and charge $ke$ sector.

The constants $V_{IJ}$ will determine the scaling dimensions of both
terms in (\ref{hv}) as well as the dimensions of the composite electron
operators $\psi_{ml}$.   In the spirit of the coupled wire model, our
strategy is to choose $V_{IJ}$ to maximize convenience.   We shall see
that  certain special choices for $V_{IJ}$ make the problem
straightforwardly solvable.

\subsection{Phases and Critical Behavior}\label{sec:III.B}

To examine the phases and critical behavior of (\ref{h00},\ref{hv}) it
is useful to introduce a variable change that separates the charged and
neutral degrees of freedom.   We define
\begin{eqnarray}
\theta_\rho &=& \phi_{1R}-\phi_{1L} + k \theta_k, \nonumber\\
\varphi_\rho &=& \varphi_k/k, \nonumber\\
\theta_\sigma &=& \phi_{1R}-\phi_{1L},\label{varphithetasigmarho}\\
\varphi_\sigma &=&(\phi_{1R}+\phi_{1L})/2 - \varphi_k/k. \nonumber
\end{eqnarray}
This transformation decouples the charge and neutral sectors
\begin{eqnarray}
\left[ \theta_\alpha(x) , \varphi_{\beta} (x') \right]  &=& 2\pi
i\delta_{\alpha\beta} \Theta(x-x') \\
\left[  \theta_\rho(x),\theta_\sigma(x')\right]  &=&
\left[\varphi_\rho(x),\varphi_\sigma(x')\right] = 0.
\end{eqnarray}
where $\alpha,\beta = \rho$ or $\sigma$.     In terms of these
operators, the elementary charge $e$ and charge $ke$ operators are given
by
\begin{eqnarray}
\psi_{ml} &=& e^{i\left(\varphi_\rho +\frac{l}{k} \theta_\rho  +
\varphi_\sigma + \frac{mk -2l}{2k}\theta_\sigma\right)}
\label{psimlvarphirho} \\
\Psi_{ML} &=& e^{i\left(k\varphi_\rho +\frac{L}{k} \theta_\rho +
{\frac{Mk-L}{r k}} \theta_\sigma\right)}.
\label{PsiMLvarphirho2}
\end{eqnarray}

It is convenient to choose the forward scattering interactions so that
in terms of these new variables the charge and neutral sectors decouple,
with no terms coupling $\theta_\rho,\varphi_\rho$ to
$\theta_\sigma,\varphi_\sigma$.   In this case,
\begin{equation}
{\cal H} = {\cal H}_\rho + {\cal H}_\sigma.
\label{h2}
\end{equation}
 The charge sector ${\cal H}^\rho$ is simply a gapless Luttinger liquid
\begin{equation}
{\cal H}_\rho =  \frac{v_\rho}{4\pi}[g_\rho(\partial_x\varphi_{\rho})^2
+ \frac{1}{g_\rho}(\partial_x\theta_{\rho})^2 ] ,
\label{hrho}
\end{equation}
while in the neutral sector
\begin{equation}
{\cal H}_\sigma =
\frac{v_\sigma}{4\pi}[g_\sigma(\partial_x\varphi_{\sigma})^2 +
\frac{1}{g_\sigma}(\partial_x\theta_{\sigma})^2 ]  + u \cos k
\varphi_\sigma + v \cos \theta_\sigma.
\label{hsigma}
\end{equation}
Since $\theta_\sigma$ and $\varphi_\sigma$ do not commute, $u$ and $v$
compete with one another.    When either term dominates, the neutral
sector is gapped.
When $v$ is large, $\theta_\sigma$ is pinned.   It
is then clear that
$\psi_{ml}$,
which involves $e^{i\varphi_\sigma}$, has a gap.   The charge sector
thus describes a ``strong clustered" Luttinger liquid of charge $ke$
particles.    On the other hand, when $u$ is large, $\varphi_\sigma$ is
pinned.   For general $m$ and $l$ the charge $e$ operator $\psi_{ml}$
involves $e^{i\theta_\sigma}$ and is gapped.   However, for $l=km/2$,
$\psi_{ml}$ does not involve $\theta_\sigma$.    Therefore this
composite charge $e$ operator does not have a gap, so the system
describes a ``weak clustered" Luttinger liquid of charge $e$ particles.

The boundary between the strong and weak clustered phases occurs when
the $u$ and $v$ terms are balanced, and is related to the critical
behavior of the $k$-state clock model.   For $u=0$, the neutral sector Hamiltonian ${\cal H}^\sigma$ is
equivalent to the sine Gordon representation of the XY model, where $v$
describes the fugacity of vortices around which the angular variable
$\varphi_\sigma$ advances by $2\pi$.  The term $u \cos k\varphi_\sigma$
then introduces a $k$-state anisotropy to the XY model, leading to a
clock model.     In this picture, $u >> v$ describes the ordered state
of the clock model, while $u<<v$ describes the disordered state.    An
equivalent dual description is to view $u$ as the fugacity for vortices
around  which  $\theta_\sigma$ advances by $k \pi$.   In this case $v
\cos \theta_\sigma$ provides the $k$-state anisotropy.

\begin{figure}
\includegraphics[width=2.5in]{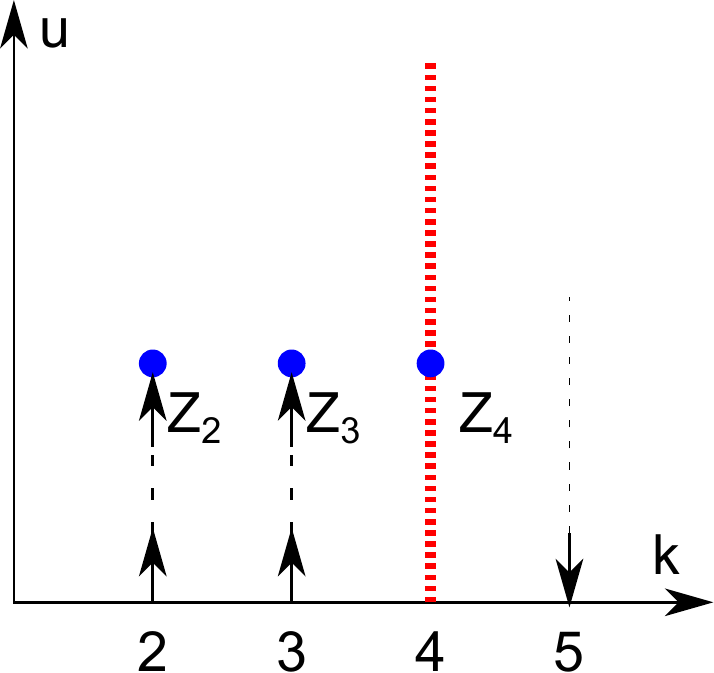}
\caption{Behavior of the $k$-state clock model at the self dual critical point
$u=v$.   For $k=2,3$ the interaction $u$ is relevant at the Luttinger liquid fixed
point, and the system flows to a $Z_k$ parafermion critical point.  For
$k>4$ the Luttinger liquid fixed point is stable.   For $k=4$ there is a
continuous line of critical points described by the orbifold CFT shown in Fig. \ref{c=1fig}.   The
$Z_4$ parafermion theory is one point on that line.  }
\label{k2345fig}
\end{figure}

The critical point occurs when the theory is self dual.   This occurs
when $u=v$ and the Luttinger parameter $g_\sigma$ is such that  the
scaling dimensions of the operators $\cos k\varphi_\sigma$ and $\cos
\theta_\sigma$ are equal to each other.    In terms of the Luttinger
parameter $g_\sigma$, the scaling dimensions are
\begin{equation}
\Delta_u = \frac{k^2}{2g_\sigma}, \quad\quad \Delta_v =
\frac{g_\sigma}{2}.
\end{equation}
The self dual point occurs at $g_\sigma = k$, where the common scaling
dimension is $\Delta = k/2$.     It follows that for $k<4$, $\Delta <2$,
so the self dual interaction $u (\cos k\varphi_\sigma + \cos
\theta_\sigma)$ is relevant.  In that case the system flows to a strong
coupling critical point describing either the Ising ($k=2$) or three
state Potts ($k=3$) transition.   For $k<4$ these are the unique
critical points with $Z_k$ symmetry, and are equivalent to the $Z_k$
parafermion critical point.    For $k>4$ we have $\Delta > 2$, and the self
dual interaction is irrelevant.  This means that in between  strong and
weak paired phase there is an intermediate phase in which the neutral
sector is gapless.    At strong coupling there may exist critical points
in a higher dimensional parameter space, such as the $Z_k$ parafermion
point, and the distinct critical point of the $k$ state Potts model.
However,  we do not have access to these strong coupling fixed points in
this theory.
The case $k=4$ is special because the interaction is marginal.    In
fact, the interaction is exactly marginal to all orders, and there
exists a line of critical points parametrized by $u$.   This fixed
line, which has continuously varying critical exponents is well known
from the study of the Ashkin-Teller model\cite{byb, ginsparg1988}, and
is described by the orbifold line of $c=1$ conformal field theory shown in Fig. \ref{c=1fig}.
Special points on this line correspond to specific distinct critical
points such as the $Z_4$ parafermion critical point
and the four state Potts model critical point.

In the following we will develop a simple description of the critical
theories on this orbifold line. This description will enable us to formulate a
coupled wire model that describes a family of non-Abelian ``orbifold"
quantum Hall states, which includes and generalizes the $Z_4$ Read
Rezayi state.    The key step that allows this progress is the
identification of a special symmetry that is present in the $k=4$
theory.

\section{$SU(2)$ symmetry for $k=4$}\label{sec:IV}

Consider the decoupled Hamiltonian in the neutral sector for $k=4$ at
the self-dual point $u=v$ and $g_\sigma = 4$.
\begin{equation}
{\cal H}_\sigma =  \frac{v_\sigma}{4\pi}[4(\partial_x\varphi_{\sigma})^2
+ \frac{1}{4}(\partial_x\theta_{\sigma})^2 ]  + u (\cos 4 \varphi_\sigma
+ \cos \theta_\sigma).
\label{huv4}
\end{equation}
This Hamiltonian still contains cosine terms, which makes analysis
beyond perturbation theory in $u$ appear difficult.   However, the
problem possesses a hidden $SU(2)$ symmetry which allows it to be cast in
a much simpler form.
We will first give a rough sketch of the $SU(2)$ symmetry which explains
why the simplification that arises.   We will then go on to discuss a
refermionization procedure that enables us to carry it out precisely.

We begin by defining chiral charge and neutral fields
\begin{eqnarray}
\phi_{\rho R} &=& \varphi_\rho + \theta_\rho/4, \nonumber \\
\phi_{\rho L} &=& \varphi_\rho - \theta_\rho/4,
\label{phirho} \\ \nonumber\\
\phi_{\sigma R} &= & \varphi_\sigma + \theta_\sigma/4,   \nonumber \\
\phi_{\sigma L} &= & \varphi_\sigma - \theta_\sigma/4,
\label{phisigma}
\end{eqnarray}
 which satisfy
\begin{equation}
\left[\phi_{\alpha A}(x),\phi_{\alpha' A'}(x') \right] =i
\frac{\pi}{2}\delta_{\alpha\alpha'} {\rm sgn} (x_A-x'_{A'})  , \\
\end{equation}
for $A = R,L$ and $\alpha = \rho,\sigma$, where we adopt the same
convention for $x_A$ as in Eq. \ref{phi1com}.  The definition of
$\phi_{\sigma A}$ is designed to simplify (\ref{huv4}) by separating the
chiral components.
The fields $\phi_{\rho A}$
 In terms of these variables we have
\begin{equation}
{\cal H}_\sigma = \frac{v_\sigma}{2\pi}[ (\partial_x\phi_{\sigma R})^2 +
(\partial_x\phi_{\sigma L})^2 ] - 2 u \cos 2\phi_{\sigma R} \cos
2\phi_{\sigma L}.
\label{hsigunrotate}
\end{equation}
The operators $\cos 2\phi_{\sigma A}$, $\sin 2\phi_{\sigma A}$ and
$\partial_x \phi_{\sigma A}$ all have dimension $\Delta = 1$.  With
appropriate numerical prefactors they define the $x$, $y$ and $z$
components of a chiral current operator $\vec J_{A=R/L}$ that satisfies
an $SU(2)_1$ current algebra.  By performing a $\pi/2$ $SU(2)$
rotation, it is possible to transform $\cos \phi_{\sigma A}$ into
$\partial_x\phi_{\sigma A}$ (up to a cutoff dependent numerical
prefactor)\cite{lgn2002}.   The Hamiltonian then takes the form
\begin{equation}
\tilde{\cal H}_\sigma = \frac{v_\sigma}{2\pi}[ (\partial_x\phi_{\sigma
R})^2 + (\partial_x\phi_{\sigma L})^2
+ 2\lambda_\sigma \partial_x\phi_{\sigma R}\partial_x\phi_{\sigma L}]
\label{hsigrotate}
\end{equation}
where $\lambda_\sigma \propto u$.   This resembles a Luttinger liquid
with a $\lambda_\sigma$ dependent Luttinger parameter.   This line of
$c=1$ fixed points parametrized by $\lambda$ defines the ``orbifold
line", which is well known in the conformal field theory
literature\cite{byb}.   It is not exactly the same as an ordinary
Luttinger liquid because the bosonic fields are compactified on an
orbifold rather than a circle, which, as will be explained further below,
modifies the operator content of the theory.

In the rotated basis, the strongly interacting theory
(\ref{hsigunrotate}) becomes a free theory (\ref{hsigrotate}), allowing
for an analysis that is nonperturbative in $u$.   In order to apply this
to quantum Hall states using the coupled wire model, the task at hand is
to learn how to describe the physical local operators (\ref{localop}) in
this rotated basis.   We have found that this is most easily
accomplished by recasting this problem in terms of a new set of $SU(2)$
fermion variables.   In addition to providing us the technical means to
accomplish the rotation, this refermionization procedure will shed light
on the relationship between the orbifold theory and the ordinary
Luttinger liquid (the circle theory).

\subsection{Fermionization}\label{sec:IV.A}

To make the $SU(2)$ symmetry present for $k=4$ explicit, we introduce
yet another basis for the four component boson field $\Phi$ defined in
(\ref{phivec}).  We define
\begin{eqnarray}
\phi_{\uparrow A} &=& \phi_{\rho A} + \phi_{\sigma A} \nonumber \\
\phi_{\downarrow A} &=& \phi_{\rho A} - \phi_{\sigma A},
\label{phiuparrowdownarrowA}
\end{eqnarray}
where $A= L$ or $R$.
These satisfy
\begin{equation}
[\phi_{sA}(x),\phi_{s'A'}(x')] = i\pi\delta_{ss'} {\rm sgn}(x_A-x'_{A'})
.
\end{equation}
where $s = \uparrow,\downarrow$ and we use the same convention for $x_A$
as Eq. \ref{phi1com}.    These operators resemble the chiral boson
operators in a theory of $SU(2)$ fermions.   This motivates us to
fermionize them by defining
\begin{equation}
\psi_{s A}(x) = \frac{e^{i\pi N_\sigma/2} }{\sqrt{2\pi x_c}} e^{i
\phi_{s A}(x) } .
\label{psiupdown}
\end{equation}
 $\psi_{s A}(x)$ obey the anticommutation relations for $SU(2)$
 fermions.
Here $x_c$ is a short distance cutoff, which is necessary to identify
the numerical prefactors.  Note that since $[\phi_{\uparrow
A},\phi_{\downarrow A'}]=0$ it is necessary to include a factor that
ensures that $\left\{\psi_{\uparrow,A},\psi_{\downarrow,A'}\right\}=0$.
This is accomplished by the $\exp(i \pi N_\sigma/2)$ term, where
\begin{equation}
N_\sigma = N_\uparrow - N_\downarrow = \int dx \partial_x(\phi_{\sigma
R} - \phi_{\sigma L})/\pi,
\label{nsigmadef}
\end{equation}
which satisfies $[N_\sigma,\phi_{\sigma A}] = i$.   Note that when
expressed in terms of the original variables, $N_\sigma = N_1$, which is
the total number of the original charge $e$ fermions.   We also define
\begin{equation}
N_\rho = N_\uparrow + N_\downarrow
\label{nrhodef}
\end{equation}
and note that $N_\rho$ and $N_\sigma$ differ by an even number, since
the parity of the total charge is the parity of the single electron
number.

Consider the $SU(2)$ current defined by
\begin{equation}
\vec J_A = \frac{1}{2} \sum_{r,s=\uparrow,\downarrow} \psi_{r A}^\dagger
\vec \sigma_{rs} \psi_{s A}.
\end{equation}
Using (\ref{psiupdown}) and (\ref{phiuparrowdownarrowA}) it follows that
\begin{eqnarray}
J_A^x &=&   \frac{1}{2\pi x_c} \cos \phi_{ \sigma A}\nonumber\\
J_A^y &=&  \frac{1}{2\pi x_c} \sin \phi_{ \sigma A}  \label{jxyz}\\
J_A^z &=& \frac{\partial_x \phi_{\sigma A}}{4\pi}. \nonumber
\end{eqnarray}

The Hamiltonian (\ref{h1}) may now be written in terms of these $SU(2)$
fermions.
\begin{equation}
{\cal H}_{\uparrow\downarrow} = -i v_F \sum_{s,A,B} \tau^z_{AB}
\psi^\dagger_{AR} \partial_x \psi_{BR} + 2\pi\lambda v_F  J^x_R J^x_L,
\label{hupdown1}
\end{equation}
where the sum is over $s=\uparrow,\downarrow$ and $A,B = R,L$ and we
identify  $\lambda = \pi x_c^2 u/v_F $.

${\cal H}^{\uparrow\downarrow} $ is an exact representation of the
Hamiltonian (\ref{h1}) for the specific choice of forward scattering
interactions that decouples ${\cal H}^\rho$ and ${\cal H}^\sigma$ in
(\ref{h2}) and sets $g_\rho = g_\sigma = 4$.    However, our theory is
not identical to ordinary $SU(2)$ fermions, even for $\lambda=0$ because
the {\it Hilbert space} on which ${\cal H}_{\uparrow\downarrow}$ acts is
not the same.   This is the origin of the difference between the
orbifold theory and the ordinary Luttinger liquid.

The reason for the difference can be seen by expressing $\phi_{sA}$ in
terms of the original variables $\Phi$ in (\ref{phivec}).   Using
(\ref{varphithetasigmarho}) and (\ref{phiuparrowdownarrowA}) we find
\begin{eqnarray}
\phi_{\uparrow,R} &=& \phi_{1R} + \theta_4,\nonumber\\
\phi_{\uparrow,L} &=& \phi_{1L} - \theta_4,\label{phiupdown} \\
\phi_{\downarrow,R} &=& \frac{1}{2}(-\phi_{1R}-\phi_{1L} + \varphi_4
+2\theta_4 ),\nonumber\\
\phi_{\downarrow,L} &=&  \frac{1}{2}( -\phi_{1R}-\phi_{1L} + \varphi_4 -
2\theta_4).
\nonumber
\end{eqnarray}
Due to the presence of the $1/2$, this is not an $SL(2,Z)$ change of
basis.   This means that unlike for ordinary $SU(2)$ fermions,  the set
of local operators is not simply given by exponentials of integer
multiples of $\phi_{sA}$ as in Eq. \ref{localop}.    This introduces two
important modifications.

The first is that not all of the states in the $SU(2)$ fermion Hilbert
space are present in our problem.   From (\ref{phiupdown}) it can be
observed that products of fermion operators will be local operators
(with integer coefficients of $\Phi_I$) if and only if the total number
of down spin operators is {\it even}.    Thus, the only states of the
$SU(2)$ fermion theory that correspond to physical states in our theory
are the states with an even number of down spin fermions.    There is a
{\it constraint} on the Hilbert space of the form
\begin{equation}
\Xi \equiv e^{i\pi N_\downarrow}  =1
\label{constraintunrot}
\end{equation}
where $N_s = \int dx \psi_{Rs}^\dagger\psi_{Rs} +
\psi_{Ls}^\dagger\psi_{Ls}$  is the total number of  spin
$s=\uparrow,\downarrow$ fermions.    Expressed in terms of charge and
neutral variables this takes the form
\begin{equation}
\Xi = \Xi_\rho \Xi_\sigma
\end{equation}
 with
\begin{eqnarray}
\Xi_\rho &=& e^{i\pi N_\rho /2} \label{xirho}\\
\Xi_\sigma &=& e^{i \pi S^z},
\end{eqnarray}
where $S^z=N_\sigma/2$.
The operator $\Xi_\sigma$ implements a $\pi$ rotation of the spin about
the $z$ axis, which takes $\varphi_{\sigma} \rightarrow \varphi_{\sigma}
+ \pi$.    Thus, the constraint effectively reduces the compactification
radius of $\varphi_\sigma$ from $2\pi$ to $\pi$ by identifying points
that differ by $\pi$.

The second difference is that there exists an operator in our theory
that is {\it not} present in the $SU(2)$ fermion theory.   Consider the
operator
\begin{equation}
e^{i\theta_4} = e^{\frac{i}{2}(\phi_{\downarrow R} - \phi_{\downarrow
L})}
\end{equation}
This is clearly a local operator in our theory, but it can not be
expressed in terms of the $SU(2)$ fermion operators $\psi_{sA}$.   In
contrast, $e^{2i\theta_4}$, as well as the other three elementary local
operators $e^{i \phi_{1R}}$, $e^{i \phi_{1L}}$ and $e^{i\varphi_4}$ can
be written locally in terms of $\psi_{sA}$.    Thus, there are two
classes of operators (and states):   those with a ``twist" and those
without.   Acting in the spin sector, $e^{i\theta_4}$ involves $e^{i
\theta_\sigma/4}$, which introduces a $\pi$ kink into $\varphi_\sigma$.
For ordinary $SU(2)$ fermions this is not allowed because
$\varphi_\sigma$ has a compactification radius of $2\pi$.   However, due
to (\ref{constraintunrot}) it is allowed, since the $\varphi_\sigma$ and
$\varphi_\sigma + \pi$ are identified.
If we write this operator in terms of the charge and spin fields it is
\begin{equation}
e^{i\theta_4} = e^{\frac{i}{4}(\phi_{\rho R}-\phi_{\rho L}) } \sigma^+_R
\sigma^-_L
\end{equation}
where we identify the chiral {\it twist operators}
\begin{equation}
\sigma^\pm_A = e^{\pm i\phi_{\sigma A}/4}
\end{equation}
These operators play a central role in the orbifold conformal field
theory.  For $\lambda=0$ it is straightforward to see that  $\sigma_A$
(which is the ``4th root" of the dimension $1$ operator
$e^{i\phi_{\sigma A}}$) has dimension $1/16$.     In addition, there
exists an ``excited" twist operator
\begin{equation}
\tau_A^\pm = e^{\mp 3 i \phi_{\sigma A}/4}
\end{equation}
with dimension $9/16$.

Finally, it is useful to express the local composite electron operators
defined in Eq. \ref{psiml}.  Using (\ref{phiupdown}) we find
\begin{equation}
\psi_{ml} =  e^{i\left(\frac{m+1}{2}\phi_{R\uparrow} - \frac{m-1}{2}
\phi_{L\uparrow}
+\frac{l-m}{2}(\phi_{R\downarrow} - \phi_{L\downarrow})\right)}.
\label{psimlboson}
\end{equation}
Recall that $m$ is odd.  When $l$ is even, $\psi_{ml}$ involves the
twist operator and can not be expressed in terms of the $SU(2)$
fermions.
In this paper we will focus exclusively on states built from operators
in which $l$ is odd, and may be expressed in terms of the $SU(2)$
fermion operators
as
 \begin{equation}
\psi_{ml} =
\psi_{R\uparrow}(\psi_{L\uparrow}^\dagger\psi_{R\uparrow})^\frac{m-1}{2}
 (\psi_{L\downarrow}^\dagger\psi_{R\downarrow})^\frac{l-m}{2}.
\label{psimlupdown}
\end{equation}
In this expression, negative powers should be understood as implying the
substitution $\psi \leftrightarrow \psi^\dagger$. In particular, note
that $\psi_{-m,-l}$ is essentially $\psi_{ml}$ with the substitution
$R\leftrightarrow L$.

\subsection{$SU(2)$ Rotation}\label{sec:IV.B}

 In order to simplify  (\ref{hupdown1}) we now implement a $90^\circ$
 $SU(2)$ rotation that converts $J^x_{A}$ to $J^z_{A}$.   Upon
 rebosonizing, this will lead to a Hamiltonian that is quadratic in the
 boson operators even when $\lambda$ is large.    Consider the canonical
 transformation
\begin{equation}
U = e^{i\pi S^y/2}
\end{equation}
where $\vec S = \int dx (\vec J_R +\vec J_L ) $.     Under this
transformation
\begin{equation}
U^\dagger (J^x_A,J^y_A,J^z_A) U= (J^z_A,J^y_A,-J^x_A)
\end{equation}
for $A=R,L$ and
\begin{equation}
U^\dagger \psi_{rA} U = \sum_{s=\uparrow,\downarrow}[ e^{i\pi
\sigma^y/4} ]_{rs} \psi_{sA}.
\end{equation}

It is now straightforward to do the rotation by performing the canonical
transformation
\begin{eqnarray}
\tilde{\cal H}_{\uparrow\downarrow} &\equiv& U^\dagger {\cal
H}^{\uparrow\downarrow} U \\
&=& -i v_F \sum_{s,A,A'} \psi^\dagger_{sA}\tau^z_{AA'} \partial_x
\psi_{sA'}  + 2\pi\lambda v_F  J^z_R J^z_L,\nonumber
\label{hupdown2}
\end{eqnarray}
Upon
 bosonizing, the Hamiltonian in the rotated basis becomes $\tilde{\cal
 H} = {\cal H}_\rho + \tilde {\cal H}_\sigma$, where ${\cal H}_\rho$ is
 given by (\ref{hrho}) with $g_\rho=4$ and ${\cal H}_\sigma$ is given by
 (\ref{hsigrotate}).   Thus we can identify $\lambda = 4\pi x_c^2
 u/v_F$.    Using (\ref{phisigma})  this can then be recast in terms of
$\theta_{\rho,\sigma}$ and $\varphi_{\rho,\sigma}$ as
\begin{equation}
\tilde {\cal H}_\sigma =   \frac{\tilde v_\sigma}{4\pi}[\tilde g_\sigma
(\partial_x\varphi_{\sigma})^2 + \frac{1}{ \tilde
g_\sigma}(\partial_x\theta_{\sigma})^2 ]
\label{tildehsigma}
\end{equation}
with $\tilde v_\sigma = v_\sigma \sqrt{1-\lambda^2}$ and
\begin{equation}\tilde g_\sigma = 4 \sqrt{(1+\lambda)/(1-\lambda)}.
\label{gbsigma}
\end{equation}

We have cast the strongly interacting Hamiltonian in a form that allows
us to take advantage of Abelian bosonization.   It now remains to
express the local electron operators in this rotated basis.
The rotated form of the single electron operators $\psi^\dagger_{sA}$
are
\begin{eqnarray}
U^\dagger \psi^\dagger_{ \uparrow A} U &=&
\frac{1}{\sqrt{2}}(\psi^\dagger_{\uparrow A} + \psi^\dagger_{\downarrow
A}) \nonumber\\
U^\dagger \psi^\dagger_{\downarrow A} U &=&
\frac{1}{\sqrt{2}}(\psi^\dagger_{\uparrow A} - \psi^\dagger_{\downarrow
A}).
\end{eqnarray}

One could construct the rotated versions of the local composite fermion
operators in Eq. \ref{psimlupdown} by taking products of many of the
above terms.   However, these will involve a sum of many terms.
Another approach is to ask what is the simplest form of composite
operators is in the rotated basis.   To this end, consider the rotated
form of the constraint operator,
\begin{equation}
\tilde\Xi = U^\dagger \Xi U = \Xi_\rho \tilde\Xi_\sigma
\label{tildexi}
\end{equation}
with $\Xi_\rho$ given in (\ref{xirho}) and
\begin{equation}
\tilde\Xi_\sigma = e^{i\pi S^x}.
\label{tildexisigma}
\end{equation}
This has the property
\begin{eqnarray}
\tilde\Xi^\dagger \psi_{\uparrow A} \tilde\Xi &=& \psi_{\downarrow A}
\label{xipsi}\\
\tilde\Xi^\dagger \psi_{\downarrow A} \tilde\Xi &=& \psi_{\uparrow A}.
\nonumber
\end{eqnarray}
It follows that any combination of fermion operators that preserves the
constraint $\tilde\Xi=1$ must be invariant under the interchange of up
and down spins.    This invites us to consider the set of local charge
$e$ operators in the rotated basis that are built from $\psi_{ml}$ in
(\ref{psimlupdown}):
\begin{equation}
\tilde\psi_{ml} = \frac{1}{2}(\tilde\psi_{ml \uparrow} +\tilde\psi_{ml
\downarrow} )
\label{psimlrotated0}
\end{equation}
with
\begin{eqnarray}
\tilde\psi_{ml \uparrow} &=&
\psi_{R\uparrow}(\psi_{\uparrow L}^\dagger\psi_{\uparrow
R})^\frac{m-1}{2}
 (\psi_{\downarrow L}^\dagger\psi_{\downarrow R})^\frac{l-m}{2}
 \nonumber\\
\tilde\psi_{ml \downarrow} &=&
\psi_{\downarrow R}(\psi_{\downarrow L}^\dagger\psi_{\downarrow
R})^\frac{m-1}{2}
 (\psi_{\uparrow L}^\dagger\psi_{\uparrow R})^\frac{l-m}{2}
\label{psimlrotated}
 \end{eqnarray}
$\tilde\psi^\dagger_{ml}$ can be related to the unrotated
$\psi^\dagger_{ml}$ by inverting the $SU(2)$ rotation.   It will be the
sum of many different terms, but each term is guaranteed to satisfy the
constraint by having an even number of down spin fermion operators.
These operators will serve as the building blocks for our coupled wire
construction in the next section.

Upon rebosonizing, using (\ref{psiupdown}), $\tilde\psi_{ml}$ may be
expressed in the charge-spin variables defined in
(\ref{phiuparrowdownarrowA},\ref{phirho},\ref{phisigma}) as
\begin{equation}
\tilde\psi_{ml} \sim e^{i(\varphi_\rho + \frac{l}{4} \theta_\rho)}
e^{i\pi N_\sigma/2} \cos\left( \varphi_\sigma + \frac{2m-l}{4}
\theta_\sigma\right).
\label{psimlbosonized}
\end{equation}

Finally we contemplate the $SU(2)$ rotation of the twist field.    The
constraint $\Xi$ in the unrotated bases identifies $\varphi_\sigma$ with
$\varphi_\sigma + \pi$, which is equivalent to a $\pi$ rotation about
$\hat z$ that takes $(J^x,J^y,J^z)$ to $(-J^x,-J^y,J^z)$.   The rotated
constraint $\tilde \Xi$ takes $(J^x,J^y,J^z)$ to $(J^x,-J^y,-J^z)$,
which is equivalent to taking $\varphi_\sigma$ to $-\varphi_\sigma$.
Thus, rather than compactifying the circle with circumference $2\pi$ to
a smaller circle of radius $\pi$, the rotated constraint compactifies
the circle to an {\it orbifold}, which is a $2\pi$ circumference circle
with $\varphi_\sigma$ and $-\varphi_\sigma$ identified.    The rotated
twist operator $\sigma^\pm_R \sigma^\mp_L$ therefore introduces a
``kink" in which $\varphi_\sigma \rightarrow -\varphi_\sigma$ on one
side.

There is no simple representation for the rotated form of the twist
operators.
However, we saw above that for $\lambda=0$ (or equivalently $\tilde
g_\sigma = 4$) the twist operators have a simple representation in the
unrotated basis, which shows that they have dimensions
$\Delta_{\sigma^\pm} = 1/16$ and $\Delta_{\tau^\pm}=9/16$.   In fact,
these dimensions are {\it independent of the orbifold radius} and remain
the same for all values of $\lambda$ (or $\tilde g_\sigma$)\cite{dijkgraaf1989,ginsparg1988,ginsparg1988b,byb}.   This is
plausible because $\tilde g_\sigma$ can be absorbed by a suitable
rescaling of $\varphi_\sigma$.   Unlike the $\varphi_\sigma \rightarrow
\varphi_\sigma +\pi$ kink,  the $\varphi_\sigma\rightarrow
-\varphi_\sigma$ kink is invariant under rescaling $\varphi_\sigma$, so
$\Delta_{\sigma^\pm}$ and $\Delta_{\tau^\pm}$ should not depend on
$\tilde g_\sigma$.

\section{Coupled Wire Model}\label{sec:V}

We now develop a theory of fractional quantum Hall states by coupling
together the wires.
We consider an array of wires parametrized by $i$ with a magnetic flux
$b$ per unit length between any pair of neighboring wires. The array is
described by the Hamiltonian
\begin{equation}
{\cal H} = \sum_i {\cal H}_{\rho,i} + {\cal H}_{\sigma,i} + {\cal
H}_{T1,i+1/2} + {\cal H}_{T4,i+1/2} + {\cal H}_{{\rm int},i+1/2}.
\end{equation}
Here, ${\cal H}_{\rho,i}$ is given by (\ref{hrho}) for each wire, and is
parametrized by the Luttinger parameter $g_\rho$ describing the
ordinary Luttinger liquid of the charge sector.  The spin part of the
Hamiltonian, ${\cal H}_{\sigma,i}$, is given  for each wire by
(\ref{hsigma}).  For $u=v$, it may be expressed  in the rotated basis as
$\tilde{\cal H}_{\sigma,i}$ by (\ref{tildehsigma}) and is characterized
by $\tilde g_\sigma$, which identifies the point on the orbifold line in
the neutral sector.   We will choose specific values for $g_\rho$ and
$\tilde g_\sigma$, which will depend on the different states that we
construct below.   For those special values of $g_\rho$ and $\tilde
g_\sigma$, the single wire factorizes into decoupled left and right
moving chiral sectors that have the same structure as the edge states of
the quantum Hall states that we will construct, so that the single wire
is like a wide quantum Hall strip.

We consider two types of tunneling terms that couple the wires.   The
term ${\cal H}_{T1,i+1/2}$ tunnels single electrons between wires $i$
and $i+1$, and is given by
\begin{equation}
{\cal H}_{T1,i+1/2} = -t_1 \psi_{1,i}^\dagger \psi_{1,i+1} e^{i b x} +
h.c.
\end{equation}
where the oscillating exponential is due to the magnetic flux per unit
length $b$ between the wires, and we set $\hbar=e=1$.  The operator
$\psi_{1,i}$ will  in general be a sum over many terms $\psi_{ml}$ in
(\ref{Psi1}), with oscillating phases $e^{i \pi l \bar\rho_e x/2}$ due
to momentum.    We consider terms $\psi_{ml,i}^\dagger \psi_{-m-l,i+1}$
and require that the oscillating factors to cancel, giving $b = \pi l
\rho_e$.   If we define the filling factor $\nu = 2\pi \bar\rho_e/b$,
then for filling factor
\begin{equation}
\nu = 2/l
\label{fillingfactor}
\end{equation}
we allow the single electron tunneling term
\begin{equation}
{\cal H}_{T1,i+1/2} = -t_1 \psi_{ml,i}^\dagger \psi_{-m-l,i+1}.
\label{ht1}
\end{equation}
As explained in Section \ref{sec:II.B}, we will focus on the case in
which the integer $l$ is odd.

In addition we consider the tunneling of clusters of electrons between
the wires,
\begin{equation}
{\cal H}_{T4,i+1/2} = -t_4 \Psi_{4,i}^\dagger \Psi_{4,i+1} e^{i 4  bx} +
h.c.
\end{equation}
The operator $\Psi_{4,i}$ will be a sum of terms $\Psi_{ML,i}$ with
phase $e^{i \pi L \bar\rho_e x/2}$.
We again require that the magnetic field term cancels the phase due to
the momentum. So given (\ref{fillingfactor}), we consider terms
$\Psi_{ML,i} \Psi^\dagger_{-M-L,i+1}$ with $L=4l$.   In addition, from
(\ref{PsiMLvarphirho2}) it can be seen that if $L=4M$ then $\Psi_{ML,i}$
only involves the charge sector.   We will assume this without loss of
generality, since other terms will be generated by combination with the
relevant $u$ term in (\ref{hsigma}).    This leads us to write
\begin{equation}
{\cal H}_{T4,i+1/2} = -t_4 \Psi_{l,4l,i}^\dagger \Psi_{l,4l,i+1}
\label{ht4}
\end{equation}

We note that by combining single-electron and cluster-tunneling we limit
ourselves to a subset of the bosonic Laughlin states that may be created
for charge-$4e$ bosons. The latter would satisfy $\nu=8/l$, but the
weakly-clustered states and the non-Abelian states formed by wires in
the critical states occur only for $\nu=2/l$.

We will also consider an extra interaction term ${\cal H}_{{\rm
int},i+1/2}$ which couples wires $i$ and $i+1$.   These terms will be
designed to ensure that $t_1$ and $t_4$ are relevant, and will be
specified below.

In the following we will show that the interactions (\ref{ht1}) and
(\ref{ht4}) define a sequence of fractional quantum Hall states
parametrized by the odd integers $l$ and $m$.    The integer $l$
specifies the character of the state in the charge sector and determines
the filling factor, while the integer $m$ characterizes the neutral
sector and specifies a sequence of non-Abelian topological states
characterized by the orbifold conformal field theories.

In order to define the bosonized theory with multiple wires it is
necessary to specify the convention for ensuring that fermions on
different wires anticommute.   We do this by defining the boson
operators on different wires to have a non zero commutator, specified by
an ordering of the wires.   For right and left moving modes on wire $i$
at position $x$ we generalize the convention in Eq. \ref{phi1com} and
define
$x_{i R} = L+ x + 2L i$ and $x_{iL} = L-x + 2L i$.   This defines a
``raster pattern" in which $... < x_{i L} < x_{i R} < x_{i+1 L} < x_{i+1
R} < ...$.   Then, for the original fermions (Eq. \ref{phi1com})
we have
\begin{equation}
[\phi_{1 i A}(x),\phi_{1 i'A'}(x') ] = i\pi {\rm sgn}(x_{i
A}-x'_{i'A'}).
\label{phi1com2}
\end{equation}
Likewise, for the $SU(2)$ fermions in Eq. (\ref{phiuparrowdownarrowA})
we write
\begin{equation}
[\phi_{i,s A}(x),\phi_{i',s'A'}(x')] = i\pi \delta_{ss'} {\rm sgn}(x_{i
A}-x'_{i'A'}).
\end{equation}
where $s,s' = \uparrow,\downarrow$.   Note that
$[\phi_{i,\uparrow,A},\phi_{i',\downarrow,A'}]=0$.   The anticommutation
between $\psi_{i,\uparrow,A}$ and $\psi_{i',\downarrow,A'}$ is taken
into account by the prefactor $\exp i\pi N_\sigma/2$ in $\psi_{i,s,A}$,
as in (\ref{psiupdown}), where now $N_\sigma$ refers to the total spin
on all of the wires.   Similar commutation relations follow for the
chiral charge and spin modes defined in (\ref{phirho},\ref{phisigma}).
For the non chiral fields $\theta_{\rho,\sigma}$ and
$\varphi_{\rho,\sigma}$ defined in (\ref{varphithetasigmarho}), as well
as the rotated versions used in (\ref{tildehsigma}) satisfy
\begin{eqnarray}
\left[ \theta_{i\alpha}(x) , \varphi_{i'\alpha'} (x') \right]  &=& 2\pi
i\delta_{\alpha\alpha'} \Theta(x_{iR}-x'_{jR})  \nonumber \\
\left[  \theta_{i\alpha}(x),\theta_{i'\alpha'}(x')\right]  &=&
\left[\varphi_{i\alpha}(x),\varphi_{i'\alpha'}(x')\right] = 0.
\end{eqnarray}

We will begin with a discussion of the charge sector.   When the
individual wires are in a strong clustered phase in which the neutral
sector is gapped, ${\cal H}_{T4,i+1/2}$ leads to an Abelian quantum Hall
state, which can be interpreted as a strong clustered Laughlin state of
charge $4e$ bosons.   When the neutral sector on each wire is in the
critical state, we will show in the following section that ${\cal
H}_{T1,i+1/2}$ leads to a sequence of non-Abelian orbifold quantum Hall
states.

\subsection{Charge sector:  Strong clustered states}\label{sec:V.A}

Here we consider the charge sector, in which the individual wires are
Luttinger liquids describe by (\ref{hrho}) and coupled by ${\cal
H}_{T4,i+1/2}$ in (\ref{ht4}).    We will first focus on the case in
which $v \gg u$ in Eq. \ref{hsigma}, so that the neutral sector has a
gap, and $\theta_\sigma$ is pinned at $0$.
 Coupling the wires in the charge sector by ${\cal H}_{T4,i+1/2}$ then
 leads to a strong clustered fractional quantum Hall state at filling
 $\nu=2/l$ that can be viewed as a Laughlin state of charge $4e$ bosons
 at filling $\nu_{4e}=1/(8l)$.   In this case, the coupled wire
 construction of this state is the same as that in Ref.
 \onlinecite{kml2002} and \onlinecite{teokane2014}.   We repeat the
 analysis here to establish our notation because we will see similar
 steps in the following section, where  the individual wires will be at
 criticality, and the quantum Hall state is modified by the neutral
 sector. Note again that the strongly clustered states that we consider are
 only a subset of the possible strongly clustered Laughlin states for
 the $4e$ bosons, restricted  by the choice $L=4 l$, and chosen since
 they form non-Abelian states at criticality.

The charge $4e$ operator in (\ref{ht4}), given in
(\ref{PsiMLvarphirho2}), has the form
\begin{equation}
\Psi_{l,4l,i} \sim e^{i(4\varphi_{\rho,i} + l \theta_{\rho,i})}.
\label{psil4li}
\end{equation}
For a general value of $g_\rho$ in (\ref{hrho}), the operator in the
exponent involves both the right and left moving chiral fields, which
are proportional to $\varphi_\rho \pm \theta_\rho/g_\rho$.   However,
for the special value
\begin{equation}
g^*_\rho = 4/l,
\label{4/l}
\end{equation}
$\Psi_{l,4l,i}$ is a purely chiral field.   At this solvable point, the
single wire factorizes into right and left moving sectors
that are equivalent to the edge states of a charge $4e$ bosonic Laughlin
state at filling $\nu_{4e} = 1/(8l)$ .    The coupling term ${\cal
H}_{T4,i+1/2}$ then describes tunneling of charge $4e$ bosons between
the edges of quantum Hall strips associated with neighboring wires.
The $1\times 1$ $K$-matrix characterizing the edge state follows from
the commutation algebra of the operator in the exponent, and is given by
\begin{equation}
K_\rho=8l.
\end{equation}
This motivates us to define
\begin{eqnarray}
\bar\phi_{i\rho,R} &=& \frac{1}{2l} (\varphi_{\rho,i} + \frac{l}{4}
\theta_{\rho,i}) \nonumber\\
\bar\phi_{i+1\rho,L} &=& \frac{1}{2l}(\varphi_{\rho,i+1} - \frac{l}{4}
\theta_{\rho,i+1}).
\label{barphirho}
\end{eqnarray}
Note that this definition of $\bar\phi_{i\rho,R/L}$ depends on $l$ and
differs from the definition of $\phi_{i\rho,R/L}$ in (\ref{phirho}).
These operators obey
\begin{equation}
\left[\bar\phi_{i\rho,A}(x),\bar\phi_{i'\rho,A'}(x')\right]  = i\pi
K_\rho^{-1} {\rm sgn}(x_{i A}-x'_{i' A'}).
\end{equation}

The Hamiltonian (\ref{hrho}) for each wire is then
\begin{equation}
{\cal H}_{\rho,i} = \frac{v_\rho K_\rho}{4\pi}\left(
(\partial_x\bar\phi_{i\rho,R})^2 +
(\partial_x\bar\phi_{i\rho,L})^2\right),
\label{hrhoiaa}
\end{equation}
and they are coupled by
\begin{equation}
{\cal H}_{T4,i+1/2} = -t_4 \cos 8l \left(\bar\phi_{i\rho,R} -
\bar\phi_{i+1\rho,L}\right)
\end{equation}

The dimension $\Delta$ of $e^{i 8l \bar\phi_{\rho,R/L}}$ at $g_\rho =
g_\rho^*$ is
\begin{equation}
\Delta^* = K_\rho/2 = 4l.
\end{equation}
Since $2\Delta^*>2$, $t_4$ will be perturbatively irrelevant in the
absence of other interactions.   However, this term can be made relevant
by adding an additional forward scattering interaction of the form
\begin{equation}
{\cal H}_{{\rm int},i+1/2} =\frac{ \lambda_\rho v_\rho K_\rho } {2\pi}
\partial_x\bar\phi_{i\rho,R}\partial_x\bar\phi_{i+1\rho,L}.
\label{hlambdarho}
\end{equation}

To describe the locking of the wires is useful to introduce yet one more
set of variables  associated with the links between wires.
\begin{eqnarray}
\bar\theta_{i+1/2,\rho} &=& \bar\phi_{i\rho,R} - \bar\phi_{i+1\rho,L}
\nonumber\\
\bar\varphi_{i+1/2,\rho} &=& K_\rho(\bar\phi_{i\rho,R} +
\bar\phi_{i+1\rho,L})/2.
\label{barthetabarphirho}
\end{eqnarray}
Then $[\theta_{i+1/2,\rho}(x),\varphi_{i+1/2,\rho}(x')] = 2\pi
i\Theta(x-x')$, and we may write $\sum_i {\cal H}_{\rho,i} = \sum_i
\bar{\cal H}_{\rho,i+1/2}$ with
\begin{equation}
\bar {\cal H}_{\rho,i+1/2}  = \frac{{\bar v}_\rho} {4\pi}\left( {\bar
g}_\rho (\partial_x\bar\varphi_{\rho,i+1/2})^2 + \frac{1} {{\bar
g}_\rho}(\partial_x\bar\theta_{\rho,i+1/2})^2\right)
\end{equation}
and
\begin{equation}
{\cal H}_{T4,i+1/2} =    - t_4 \cos 8l \bar\theta_{\rho,i+1/2}.
\label{ht4rho}
\end{equation}
Here $\bar g_\rho =
2K_\rho^{-1}\sqrt{(1+\lambda_\rho)/(1-\lambda_\rho)}$ and $\bar v_\rho =
v_\rho \sqrt{1-\lambda_\rho^2}$.  This theory has the structure of a
sine-Gordon model, or equivalently a 2D XY model.   The $t_4$ term  is
relevant for $K^2 g_\rho < 2$.
Moreover, if $t_4$ starts large, then $g_4$ is renormalized downward,
making $t_4$ more relevant and leading to a gapped phase. The limits of
small and large initial values of $t_4$ are separated by a Kosterlitz
Thouless transition.

When $t_4$ flows to strong coupling $\theta_{\rho,i+1/2}$ is locked in
one of the minima of the cosine. Since  $\theta_{\rho,i+1/2}$ is an
angular variable, defined modulo $2\pi$,  there are $K = 8l$ distinct
minima of the cosine,
\begin{equation}
\theta_{\rho,i+1/2}^* =  \frac{\pi}{4l} Q
\label{thetarho*}
\end{equation}
where $Q$ is an integer mod $8l$.
A kink in which $\bar\theta_{\rho,i+1/2}$ advances by $\pi/(4l)$
corresponds to an elementary Laughlin quasiparticle of charge $e^* =
e/(2l)$.

In the following section we will consider the case in which $u=v$ in
(\ref{hsigma}) so that the neutral sector is gapless on each wire.   In
this case, the interaction in the charge sector still opens up a charge
gap, and most of the analysis of this section remains valid.    However
additional tunneling terms coupling the wires will be necessary to open
a gap in the neutral sector.  We will see that quasiparticles in the
charge sector are then bound to neutral sector excitations described by
primary fields of the orbifold theory.

\subsection{Neutral sector: Orbifold states} \label{sec:V.B}
We now construct the orbifold quantum Hall states
\cite{BarkeshliWen2011}.
We consider charge $e$ tunneling between wires (\ref{ht1}) in the case
where the individual wires are in the critical state.    We will assume
that $g_\rho = 4/l$ and that the charge sector is gapped by $t_4$ due to
(\ref{ht4rho}), so that $\theta_{\rho,i+1/2}$ is pinned, and given by
(\ref{thetarho*}).   We will work in the rotated basis for the neutral
sector, in which $\tilde {\cal H}_\sigma$ in (\ref{tildehsigma})
describes the orbifold line, parametrized by $\tilde g_\sigma$.    The
charge $e$ tunneling term $H_{T1,i+1/2}$ involves the rotated charge $e$
operators $\tilde \psi_{ml}$ and $\tilde \psi_{-m-l}$. Using
(\ref{psimlbosonized}) we write these as
\begin{eqnarray}
\tilde\psi_{ml}\equiv \tilde\psi_R &\sim& e^{i(\varphi_\rho +
\frac{l}{4}\theta_\rho)} e^{i\pi N_\sigma/2} \cos(\varphi_\sigma +
\frac{p}{4}\theta_\sigma), \nonumber\\
\tilde\psi_{-m-l}\equiv \tilde\psi_L &\sim& e^{i(\varphi_\rho -
\frac{l}{4}\theta_\rho)} e^{i\pi N_\sigma/2} \cos(\varphi_\sigma -
\frac{p}{4}\theta_\sigma).  \label{tildepsiml}
\end{eqnarray}
Here we have introduced a new odd integer
\begin{equation}
p = 2m-l.     \label{pml relation}
\end{equation}
It will be useful to consider the odd integers
$p$ and $m$ to be the independent parameters.    In this case,
(\ref{fillingfactor}) becomes
\begin{equation}
\nu = \frac{2} {2m -p}
\end{equation}
We will suppose for simplicity that $p$ is positive.    This will define
the direction of propagation of the neutral sector edge modes. The
number     $l = 2m-p$ (and hence $\nu$) can then be positive or
negative,
which specifies whether the neutral and charge edge modes propagate in
the same direction or in opposite directions.

 In the previous section we introduced the chiral operators
 $\bar\phi_{i\rho,R/L}$ in (\ref{barphirho}) ($R$ and $L$ will be
 interchanged if $l<0$).    We now introduce corresponding operators for
 the neutral sector,
\begin{eqnarray}
\bar\phi_{i\sigma,R} &=& \frac{1}{p}(\varphi_{\sigma,i} +
\frac{p}{4}\theta_{\sigma,i}), \nonumber\\
\bar\phi_{i+1\sigma,L} &=& \frac{1}{p}(\varphi_{\sigma,i+1} -
\frac{p}{4} \theta_{\sigma,i+1}).
\label{barphisigma}
\end{eqnarray}
As in the previous section, these operators are not in general purely
chiral.    However, for a particular choice of $\tilde g_\sigma$ in
(\ref{tildehsigma}),
\begin{equation}
\tilde g_\sigma^* = 4/p,
\label{4/p}
\end{equation}
the chiral fields decouple, and
\begin{equation}
\tilde{\cal H}_{\sigma,i+1/2} = \frac{\tilde v_\sigma K_\sigma}{4\pi}
\left ((\partial_x\bar\phi_{i\sigma,R})^2 +
(\partial_x\bar\phi_{i\sigma,R})^2 \right)
\end{equation}
with
\begin{equation}
K_\sigma = 2p.
\end{equation}
The chiral fields satisfy
\begin{equation}
\left[\bar\phi_{i\sigma,A}(x),\bar\phi_{i'\sigma,A'}(x')\right]  =  i\pi
K_\sigma^{-1} {\rm sgn}(x_{iA}-x'_{i' A'}).
\end{equation}

For different values of $p$, the value of $g_\sigma=g_\sigma^*$ puts the
theory for a single wire at  specific points on the orbifold line that
form a set of known rational conformal field theories.
An ordinary Luttinger liquid with $K_\sigma = 2p$ defines a conformal
field theory compactified on a circle of radius $R_{\rm circle}=\sqrt{p/2}$\cite{Note1}.   In the
present case, the constraint $\tilde \Xi$, which relates
$\varphi_\sigma$ to $-\varphi_\sigma$ defines the theory on an orbifold
of the same radius.
We will defer the discussion of the operator content of these theories
to the next session.   Here we will note that $p=1$, with $g_\sigma = 4$
corresponds to either $\lambda_\sigma = 0$ in (\ref{hsigrotate}) or
$u=0$ in (\ref{hsigunrotate}).    Since $u=0$, this state can be
described either in the rotated or the unrotated basis.   It is an
Abelian state, where the neutral sector is described equivalently as a
$R_{\rm orbifold}=1/\sqrt{2}$ orbifold or an ordinary Luttinger liquid (circle) with
$R_{\rm circle}=\sqrt{2}$. These describe the $K=8$ (or $U(1)_8$) state.   The $p=1$
state occurs at filling factors
\begin{equation}
\nu = \frac{2}{2m - 1} = ..., -2/7, -2/3, 2, 2/5, ...
\label{p=1filling}
\end{equation}
where we recall that negative filling factors imply states with
counter-propagating charge and neutral modes.

 The value $p=3$  defines a sequence of quantum Hall states at filling
\begin{equation}
\nu = \frac{2} {2m - 3} = ..., -2/5, -2, 2/3, 2/7, ...
\label{p=3filling}
\end{equation}
These correspond to the filling factors of the $k=4$ sequence of Read
Rezayi states, including conjugate states with counter-propagating charge
and neutral modes.  This value of   $p=3$ corresponds to the
$R=\sqrt{3/2}$ orbifold which is precisely the $Z_4$ parafermion point
on the orbifold line\cite{ginsparg1988b,ginsparg1988}.

Higher values of $p = 5, 7, ...$ correspond to a generalization of the
$k=4$ Read Rezayi states.    There exists a distinct quantum Hall state
for each odd integer value of $p$.   These fall into two categories:
the states with $p=1$ mod $4$ are defined at the filling factors in
(\ref{p=1filling}), while the states with $p=3$ mod $4$ occur at filling
factors in (\ref{p=3filling}).

The electron operators $\tilde\psi^\dagger_{ml}$ have a dimension that
is a sum of pieces due to the charge and neutral sectors: $\Delta =
\Delta_\rho + \Delta_\sigma$.   We find
\begin{equation}
\Delta_\rho = |l|/4, \quad   \Delta_\sigma = p/4.
\end{equation}
Note that for $p=3$, $\Delta_\sigma$ matches the dimension $3/4$ of the
$Z_4$ parafermion operator in the $Z_4$ fermion conformal field theory.
This connection will be discussed further in the following section.

The charge $e$ tunneling operator will have dimension $2\Delta =
(|l|+p)/2$.   Therefore, if $|l|+p > 4$ then $t_1$ will be irrelevant in
the absence of other interactions.   As in the previous section, adding
a term (\ref{hlambdarho}) in the charge sector can reduce $\Delta_\rho$.
However, the situation is more complicated in the neutral sector because
the operator analogous to (\ref{hlambdarho}), proportional to
$\partial_x\bar\phi_{i\sigma,R}\partial_x\bar\phi_{i+1\sigma,L}$ is not
an allowed local operator in the theory.    Once the charge sector is
repaired, the condition becomes $|p|>4$, so for $p=1,3$ there is no
issue.   But for $|p| = 5, 7, ...$ an additional interaction in the
neutral sector is required to make $t_1$ relevant.    In fact there {\it
is} an additional interaction that can be added in the neutral sector
that can make $t_1$ relevant for the entire sequence of orbifold states.
We will explain this problem and its solution in Appendix
\ref{appendix:A}.   For now, we will simply assume that $t_1$ is
relevant and explore the properties of the resulting strong coupling
state.

We now consider the coupling between the wires generated by ${\cal
H}_{T1,i+1/2}$ in (\ref{ht1}).
We  again define variables analogous to (\ref{barthetabarphirho})
associated with the links between wires.   Due to the symmetry relating
$\phi \rightarrow -\phi$ it is useful to treat $\bar\theta$ and
$\bar\varphi$ symmetrically.   We define
\begin{eqnarray}
\bar\theta_{i+1/2,\sigma} &=& \bar\phi_{i\sigma,R} -
\bar\phi_{i+1\sigma,L} \nonumber\\
\bar\varphi_{i+1/2,\sigma} &=& \bar\phi_{i\sigma,R} +
\bar\phi_{i+1\sigma,L}\label{barthetaphivarsigma}.
\end{eqnarray}
These obey
\begin{equation}
[\bar\theta_{i+1/2,\sigma}(x),\bar\varphi_{i+1/2,\sigma}(x')] =
\frac{2\pi i}{p} \Theta(x-x')
\end{equation}

Evaluation of ${\cal H}_{T1,i+1/2}$ requires a careful treatment of the
commutation relations between the fields when combining exponentials.
Consider the term $
\tilde\psi^\dagger_{i,R} \tilde\psi_{i+1,L}$, which using
(\ref{tildepsiml},\ref{barphirho},\ref{barphisigma}) can be written in
the form
\begin{equation}
e^{-2i l \bar\phi_{i\rho,R}} e^{2i l \bar\phi_{i+1\rho,L}} \cos p
\bar\phi_{i\sigma,R} \cos p \bar\phi_{i+1\sigma,L}.
\end{equation}
Using the fact that  $[\bar\phi_{i,\rho R},\bar\phi_{i+1,\rho' L}] = -
i\pi/(8l)$ it follows that $e^{-i 2 l \phi_{i,\rho R}} e^{i 2 l
\bar\phi_{i+1 \rho L}} = e^{-i l \pi/4} e^{-i 2l \theta_{i+1/2,\rho}}$.
Using similar considerations for $\bar\phi_{i,\sigma,R/L}$, we find
(suppressing the $i+1/2$ subscripts for brevity)
\begin{equation}
\tilde\psi^\dagger_{i,R} \tilde\psi_{i+1,L} \sim e^{-i 2l
(\bar\theta_\rho - \bar\theta_0)}(  \cos p\bar\theta_{\sigma} - i s_p
\cos p\bar\varphi_{\sigma}),
\label{psiproduct}
\end{equation}
where $\bar\theta_0 = -\pi (p+l)/(8l)$ and
\begin{equation}
s_p = e^{i \pi (p+1)/2} = \left\{\begin{array}{ll} -1  & p=1 \ {\rm
mod}\ 4 \\ +1 & p=3 \ {\rm mod} \ 4.\end{array}\right.
\label{sp}
\end{equation}
It is convenient to absorb the unimportant constant $\bar\theta_0$ into
$\bar\theta_\rho$ by replacing $\bar\theta_\rho-\bar\theta_0 \rightarrow
\bar\theta_\rho$.     The electron tunneling term connecting neighboring
wires then takes the form
\begin{equation}
{\cal H}_{1T} = -4 t_1 \left(\cos 2 l \bar\theta_\rho \cos
p\bar\theta_\sigma -s_p \sin 2 l \bar\theta_\rho \cos
p\bar\varphi_\sigma \right).
\label{ht1thetarhophisigma}
\end{equation}

The form of (\ref{ht1thetarhophisigma}) differs from that of
(\ref{ht4rho}) because it is the sum of two cosine terms that involve
the non commuting operators $\theta_\sigma$ and $\varphi_\sigma$.
Ordinarily, such operators would compete with one another because they
can not be simultaneously pinned.   But for (\ref{ht1thetarhophisigma})
the $\pi/2$ phase shift between $\cos 2l\bar\theta_\rho$ and $\sin 2l
\bar\theta_\rho$ plays an essential role.  Recall that ${\cal
H}_{T2,i+1/2}$ pins $\bar\theta_\rho$ at an integer multiple of
$\pi/(4l)$.      For a given minimum only one of the terms in
(\ref{ht1thetarhophisigma}) is operative.    Specifically, for
$\bar\theta_\rho = Q \pi/(4l)$ we have
\begin{equation}
{\cal H}_{1T} = - 4 t_1 \left\{\begin{array}{ll}
 \cos(p\bar\theta_\sigma)  & {\rm for} \  Q = 0 \ {\rm mod}\ 4, \\
s_p \cos(p\bar\varphi_\sigma) & {\rm for} \ Q = 1\ {\rm mod}\ 4, \\
 -\cos(p\bar\theta_\sigma)  & {\rm for} \  Q = 2 \ {\rm mod}\ 4, \\
- s_p \cos(p\bar\varphi_\sigma) & {\rm for} \ Q = 3 \ {\rm mod}\ 4.
\end{array}\right.  \label{pinsigma}
\end{equation}
Thus, depending on the parity of $Q$, {\it either} $\bar\theta_\sigma$
{\it or} $\bar\varphi_\sigma$ is pinned, resulting in a bulk energy gap
in the neutral sector.  Combined with the
gap provided by   (\ref{ht4rho})  in the charge sector, this results in
a quantum Hall state with a complete bulk energy gap.

\section{Quasiparticles in the Orbifold States}\label{sec:VI}

In this section we consider the structure of the quasiparticle
excitations in the orbifold states.   There are two ways to analyze the
quasiparticles.   The first is to characterize the 1+1D conformal field
theory describing the edge states\cite{mooreread1991,wenbook}.   This
can be done by considering the theory of a single wire at the solvable
point where the right and left moving chiral sectors decouple.    In
general, the edge states are characterized by an Abelian charge sector
described by a $c=1$ bosonic charge mode compactified on a circle with
radius determined by $l$ in (\ref{4/l}), along with a non-Abelian
neutral sector characterized by a $c=1$ neutral bosonic mode
compactified on an orbifold with radius defined by $p$ in (\ref{4/p}).
The physical quasiparticle operators then involve specific combinations
of the charge and neutral primary fields.   We will see that our Abelian
bosonization approach to describing the orbifold sector allows a simple
representation of these operators which then allows a straightforward
determination of the conformal dimension of the quasiparticle operators.

A second approach to understanding the quasiparticles is to consider
topological field theory characterizing the 2+1D bulk\cite{bais2009}.
Bulk quasiparticles, which exist on the links between the wires, are
described by kinks in $\bar\theta_\rho$ and $\bar\theta_\sigma$ or
$\bar\varphi_\sigma$ defined in
(\ref{barthetabarphirho},\ref{barthetaphivarsigma}).     Again, we will
see that our Abelian bosonization approach allows an understanding of
the non-Abelian braiding properties of these quasiparticles.    We will
present a simple construction that allows us to determine topological
S-matrix, which combined with the conformal dimensions of the
quasiparticles completely characterizes the non-Abelian braiding
properties of the state.

\subsection{Edge state theory}\label{sec:VI.A}

At the edge, the Hamiltonian is described by a chiral theory of the form
\begin{equation}
H_A^{\rm edge} =  \frac{\tilde v_\rho K_\rho}{4\pi}
(\partial_x\bar\phi_{A\rho})^2 + \frac{\tilde v_\sigma K_\sigma}{4\pi}
(\partial_x\bar\phi_{A\sigma})^2
\label{hedgea}
\end{equation}
where $A=R/L$ specifies the right and left moving sectors.   The
operator content of the edge theory can be determined by considering the
set of local operators that couple only to the edge states.

There are two classes of operators:  (1) local operators which act on a
single chiral edge.   These are ``trivial" operators, which describe the
creation of integer
charges in the edge states.  (2)  quasiparticle operators.    These are
operators which can not be written locally on a single chiral edge, but
a local operator can describe tunneling from
one edge to the other.    The possible quasiparticle backscattering
terms can be constructed by considering the set of charge neutral local
operators that couple the right and left
moving sectors.    In general, operators in either class will be a
product of an operator in the charge sector, and an operator in the
neutral sector.   The charge sector operator
determines the charge of the quasiparticle, and is the same as one of
the quasiparticle operators in the strong clustered state.    The
neutral sector operators will be described by primary fields
of the orbifold conformal field theory.   The structure of these
operators is well known in the conformal field theory
literature\cite{dijkgraaf1989}.
Our formulation provides an explicit bosonized representation for some
of these operators that allows many properties to be simply understood.

We will begin with a discussion of the local operators.    We will then
discuss two classes of quasiparticles: quasiparticles without a twist
and quasiparticles with a twist.

\subsubsection{Local Electron operators}\label{sec:VI.A.1}

Local operators can be built out of powers of the charge $e$ electron
operators $\tilde\psi_{R,L}$ defined in
(\ref{psimlrotated0},\ref{psimlbosonized}), which act on a single chiral
sector.   The charge $e$ operator may be factored into charge and
neutral sector components as
\begin{equation}
\tilde\psi_{R} = e^{i2l \bar\phi_{\rho R}} \Psi_{+,R}.
\end{equation}
The neutral part is
\begin{equation}
\Psi_{+,R} = e^{i\pi N_\sigma/2} \cos  p \bar\phi_{\sigma R}
\end{equation}
We also define
\begin{eqnarray}
\Psi_{-,R} &=& \Psi_{+,R}^\dagger  \nonumber \\
&=&  \cos  p \bar\phi_{\sigma R} e^{-i\pi N_\sigma/2} \\
&=&  -s_p e^{-i\pi N_\sigma/2} \sin p \bar\phi_{\sigma R},  \nonumber
\end{eqnarray}
where $s_p$ is defined in (\ref{sp}). The operator
$\Psi_{-,R}$ fits into the charge $-e$ operator $\tilde\psi_R^\dagger$.
Similarly, a charge $2e$ operator can be written
\begin{equation}
\psi_{2e,R} = e^{i4l \bar\phi_{\rho R}} \Theta_R.
\label{psi2er}
\end{equation}
The form of $\Theta$ can be deduced by forming an operator product of
two $\Psi_+$ operators.   By keeping track of the commutators involving
$\exp(i\pi N_\sigma/2)$ this is found to be
\begin{equation}
\Theta_R = e^{i\pi N_\sigma} \partial_x\bar\phi_{\sigma R}.
\label{thetaop}
\end{equation}

In general, a charge $N e$ local operator can be written as $\exp i 2lN
\phi_{\rho,R}$ times a neutral sector operator, which depending on $N$
{\rm mod} 4 is one of $(1, \Psi_{+,R},\Theta_R,\Psi_{-,R})$.

The operators $(1 ,\Psi_+,\Theta,\Psi_{-})$, which form a $Z_4$ fusion
algebra, are a subset of the primary fields of the orbifold theory for
odd integer $p$.
It is straightforward to check that their conformal dimensions $\Delta$
are
\begin{equation}
\Delta_{\Psi_\pm} = p/4;  \quad\quad
\Delta_\Theta = 1.
\end{equation}
For $p=3$, which is the $Z_4$ parafermion point of the orbifold line,
the operators $\Psi_\pm$ and $\Theta$ are precisely the bosonized
representation the $Z_4$ parafermion operators discussed in Ref.
\onlinecite{lgn2002}.   In the present approach, this bosonized form
emerges naturally from the coupled wire model, and is generalized to any
odd value of $p$.

There exists an additional local charge neutral operator of the form
\begin{equation}
\Lambda_R \equiv \tilde\psi_{ml,\uparrow}^\dagger
\tilde\psi_{ml\downarrow} + \tilde\psi_{ml,\downarrow}^\dagger
\tilde\psi_{ml\uparrow}
 \sim \cos 2p \bar\phi_{\sigma R}.
\end{equation}
This operator appears non trivial in the neutral sector.   However,
since it is local (invariant under $\uparrow \leftrightarrow
\downarrow$) and charge neutral, it is allowed to appear in the
Hamiltonian.   This dimension $p$ operator should therefore be
considered a descendant of the trivial operator.   Combining this
operator with $\Psi_{\pm R}$ or $\Theta_R$ yields descendants of those
operators.   For instance $\Theta_R \times \Lambda_R  \sim \sin 2p
\bar\phi_{\sigma R}$ is a descendant of $\Theta$ with dimension $p$.

\subsubsection{Quasiparticle Operators}\label{sec:VI.A.2}

In addition to the local charge $e$ excitations, there are additional
fractionally charged quasiparticles.   These can not be created locally,
but they can tunnel from one edge to another via a local operator.    In
the coupled wire model, these quasiparticle tunneling processes
are given by local backscattering terms on a single wire.
A general local charge 0 operator can be written in terms of our
original bosonic fields in (\ref{phivec})
as
\begin{equation}
{\cal V} = e^{i{\bf N}\cdot\Phi},
\end{equation}
 where the integer valued vector ${\bf N}$ satisfies ${\bf N}\cdot {\bf
 t}=0$ for ${\bf t} = (1,1,0,4)$.   Such an operator can be factored
 into its charge and neutral components, and will have the form
\begin{equation}
{\cal V} = {\cal O}^\rho {\cal O}^\sigma = e^{i Q (\phi_{\rho R} -
\phi_{\rho L})} {\cal O}^\sigma.
\end{equation}
Such an operator describes the backscattering of a charge $q e^* = Q
e/(2l)$ quasiparticle.   In general, such a quasiparticle involves an
operator in the neutral sector.
The distinct operators ${\cal O}_\sigma$ in the neutral sector will be
identified with the primary fields of the orbifold CFT.   We will first
summarize the quasiparticle types of the orbifold states in terms of the
known primary fields of the orbifold CFT\cite{dijkgraaf1989}.   We will
then show how those quasiparticle operators arise in our bosonized
theory.

\begin{table}
\begin{tabular}{c|c|c|l}
{\rm Operator} &  Ref. \onlinecite{dijkgraaf1989}  &  {\rm Dimension}  &
{\rm Bosonized representation} \\
\hline
$1$  & $1$ &  $0$    &${\cal O}_1\   \sim   1$\\
$\Psi_{+}$   & $\phi^1_N $ &   $p/4$  &   ${\cal O}_{\Psi_+}  \sim
\cos(p\bar\phi_{R\sigma}) \cos (p \bar\phi_{L\sigma} )$  \\
$\Psi_{-}$   & $\phi^2_N $ &   $p/4$  &   ${\cal O}_{\Psi_-}  \sim
\sin(p\bar\phi_{R\sigma}) \sin (p \bar\phi_{L\sigma} )$    \\
$\Theta$  &  $j$ & $1$ &  ${\cal O}_\Theta \ \sim   \partial_x
\bar\phi_{R\sigma}\partial_x \bar\phi_{L\sigma}$ \\
$\Phi_\lambda$ & $\phi_\lambda$ &  $\lambda^2/4p $  &  ${\cal
O}_{\Phi_\lambda} \sim \cos(\lambda(\bar\phi_{R\sigma}
-\bar\phi_{L\sigma})) $\\
 $\sigma^\pm$ & $\sigma_{1,2}$ & $ 1/16$ &  ${\cal O}_{\sigma^\pm}$  \\
$\tau^\pm  $ & $\tau_{1,2}$ & $ 9/16 $  &  ${\cal O}_{\tau^\pm} \sim
{\cal O}_{\sigma^\pm} {\cal O}_\Theta $
\end{tabular}

\caption{Primary fields of the orbifold conformal field theory.   All of
the operators except the twist fields have a simple bosonized
representation.}
\label{primariestable}
\end{table}

The primary fields of the orbifold theory for odd integer $p$ are
summarized in Table \ref{primariestable}.   There are three classes of
fields, which include   (1)  the $Z_4$ fields $\Psi_\pm$ and $\Theta$
introduced above, (2)
a set of $p-1$  ``fractional" fields $\Phi_\lambda$, with dimension
$\lambda^2/4p$.
and (3) a set of four twist fields, which includes $\sigma^\pm$ with dimension $1/16$
and $\tau^\pm$ with dimension $9/16$.   The neutral sector operator $a$
associated with a given quasiparticle depends on the quasiparticle
charge $Q$ mod 4.   We find
\begin{eqnarray}
Q = 0 \ {\rm mod}\ 4  \quad &a& = 1, \Theta, \Phi_{\lambda = {\rm even}}
\label{Q=0} \nonumber \\
Q = 1 \ {\rm mod}\ 4 \quad &a& = \sigma^+, \tau^+ \label{Q=1}\\
Q = 2 \ {\rm mod}\ 4 \quad &a& = \Psi_\pm, \Phi_{\lambda = {\rm odd}}
\label{Q=2} \nonumber \\
Q = 3 \ {\rm mod}\ 4 \quad &a& = \sigma^-, \tau^-.  \label{Q=3}
\nonumber
\end{eqnarray}
The distinct quasiparticle types are defined for charges $0 \le Q < 2l$,
which leads to a total of $l(p+7)/2$ quasiparticle types.

By examining the possible local charge neutral operators on a single
wire, we now identify the local operators that backscatter these
quasiparticles, and identify the explicit form of the operators of the
orbifold theory.   When expressed in terms of the $SU(2)$ fermions in
the rotated basis, local operators are invariant under the interchange
of $\uparrow$ and $\downarrow$ spins.

We first consider the quasiparticles that are trivial in the neutral
sector.   Consider the local operator (in the rotated basis)
\begin{equation}
{\cal V}_0 = (\psi_{R\uparrow}^\dagger
\psi_{R\downarrow}^\dagger)^\frac{Q}{4}
(\psi_{L\uparrow} \psi_{L\downarrow})^\frac{Q}{4} \sim e^{i  Q
(\bar\phi_{R\rho}-\bar\phi_{L\rho})}
\label{calv0}
\end{equation}
where $Q$ is an integer multiple of $4$ identified with (\ref{Q=0}).
This operator tunnels a charge $Q e/(2l)$ quasiparticle.  Since it does
not involve $\bar\phi_\sigma$, such a quasiparticle can be combined with
any other quasiparticle without changing its topological class in the
neutral sector.   It follows that the allowed topological classes for
quasiparticles depends on $Q$ mod $4$.

Consider next a neutral $Q=0$ quasiparticle describing the $\Theta$
field.    The operator
\begin{equation}
{\cal V}_1 = (\psi^\dagger_{R \uparrow} \psi_{R
\uparrow}-\psi^\dagger_{R \downarrow} \psi_{R
\downarrow})(\psi^\dagger_{L \uparrow} \psi_{L \uparrow}-\psi^\dagger_{L
\downarrow} \psi_{L \downarrow}).
\end{equation}
is local in the rotated basis.
While the individual terms in the product are not invariant under
$\uparrow\leftrightarrow\downarrow$, the product is invariant.
Expressed in the bosonized variables this has the form
\begin{equation}
{\cal V}_1 \equiv {\cal O}_\Theta \sim \partial_x\bar\phi_{R\sigma}
\partial_x \bar\phi_{L\sigma} \sim \Theta_R \Theta_L
\label{calv1}
\end{equation}
This can be interpreted as an operator that tunnels a neutral
quasiparticle
from one edge to the other.   Importantly, $\Theta$ by itself {\it not}
a local operator, so the neutral $\Theta$ quasiparticle is distinct from
the identity.

Next consider the local operator
\begin{equation}
{\cal V}_2 = \psi_{R\uparrow}^\dagger \psi_{L\downarrow} +
\psi_{R\downarrow}^\dagger \psi_{L\uparrow}.
\end{equation}
This has the bosonized form
\begin{equation}
{\cal V}_2 =  e^{i 2 (\bar\phi_{R\rho}-\bar\phi_{L\rho})}  \cos  p  (
\bar\phi_{R\sigma}+\bar\phi_{L\sigma}).
\end{equation}
Using the fact that $\cos  p  ( \bar\phi_{R\sigma}+\bar\phi_{L\sigma}) =
e^{i p \pi/4} (e^{i p \bar\phi_{R\sigma}} e^{i p \bar\phi_{L\sigma}}
+e^{-i p \bar\phi_{R\sigma}} e^{-i p \bar\phi_{L\sigma}} ) $,
this can be written
\begin{equation}
{\cal V}_2 = e^{i 2  (\bar\phi_{R\rho}-\bar\phi_{L\rho}) + i p\pi/4}
\left( \Psi_{+,R}^\dagger \Psi_{+,L} - \Psi_{-,R}^\dagger \Psi_{-,L}
\right).
\label{o2psi}
\end{equation}
Though this does not have a factorized form,  we will see below that
there exists another local operator
\begin{equation}
{\cal V}_3 =  e^{i 2 (\bar\phi_{R\rho}-\bar\phi_{L\rho})}  \cos p  (
\bar\phi_{R\sigma}-\bar\phi_{L\sigma}).  \label{o3}
\end{equation}
${\cal V}_3$ has a similar form as (\ref{o2psi}), except with a plus
sign:
\begin{equation}
{\cal V}_3 = e^{i 2  (\bar\phi_{R\rho}-\bar\phi_{L\rho})+i p \pi/4}
\left( \Psi_{+,R}^\dagger \Psi_{+,L} +\Psi_{-,R}^\dagger\Psi_{-,L}
\right).
\label{o3psi}
\end{equation}
It follows that the combination of the two defines a local tunneling
term for a $Q=2$ (charge $2e^*= e/l$) quasiparticle, with local
tunneling operators
\begin{equation}
{\cal O}_{\Psi_\pm} =  \Psi_{\pm R}^\dagger \Psi_{\pm L}.
\end{equation}
Due to the existence of the neutral non-trivial quasi-particle $
\Theta$ there are two distinct quasi-particles of this type for each charge $2e^*q$ and $0<q\le l$.

We next identify the class of quasiparticles associated with the
fractional fields $\Phi_\lambda$.   Consider a charge neutral local
operator of the form
\begin{eqnarray}
{\cal V}_4^{Q,n} &= (\psi^\dagger_{R\uparrow})^{\frac{Q}{2}+n}
(\psi_{R\downarrow})^n (\psi^\dagger_{L\downarrow})^n
(\psi_{L\uparrow})^{\frac{Q}{2}+n}\nonumber\\  +&
(\psi^\dagger_{R\downarrow})^{\frac{Q}{2}+n} (\psi_{R\uparrow})^n
(\psi^\dagger_{L\uparrow})^n (\psi_{L\downarrow})^{\frac{Q}{2}+n}
\label{o4}
\end{eqnarray}
where $Q$ is an even integer, to be identified with
(\ref{Q=0},\ref{Q=2}).  In boson variables this has the form,
\begin{equation}
{\cal V}_4^{Q,n} =  e^{i  Q (\bar\phi_{R\rho}-\bar\phi_{L\rho})}
\cos (\frac{Q}{2}+2n)(\bar\phi_{R\sigma}-\bar\phi_{L\sigma})\\
\end{equation}
We will identify these operators with the tunneling of charge $Q e^* = Q
e/(2l)$ quasiparticle associated with the primary fields $\Phi_\lambda$
of the orbifold theory, given by
\begin{equation}
{\cal O}_{\Phi_\lambda} = \cos \lambda (
\bar\phi_{R\sigma}-\bar\phi_{L\sigma}).
\label{philambda}
\end{equation}
For $Q=0$ mod $4$ $\lambda$ will be even, while for $Q=2$ mod $4$
$\lambda$ will be odd.   It can be seen that $\Phi_{-\lambda} =
\Phi_\lambda$, so only positive values of $\lambda$ are independent.
Moreover, for $\lambda = p$, ${\cal O}_{\Phi_p}$ is the operator
promised in (\ref{o3}), which
is a combination of ${\cal O}_{\Psi_\pm}$.  Thus, there are $p-1$
independent values $\lambda = 1, 2, ..., p-1$.

The operators ${\cal O}_{\Phi_\lambda}$ can not be factored into a
product of right and left moving operators.    Our inability to
factorize this operator is an indication that $\Phi_\lambda$ is a
non-Abelian quasiparticle with multiple fusion channels.   Nonetheless,
our bosonized representation of the tunneling operator allows us to
understand properties of the $\Phi_\lambda$ operators in the orbifold
CFT.   The dimension of $\Phi_\lambda$ follows from (\ref{philambda}),
and is given by
$\Delta(\Phi_\lambda) = \lambda^2/4p$.
Considering a product of operators ${\cal O}_\lambda \times {\cal
O}_{\lambda'}$ we can conclude that $\Phi_\lambda$ obeys a non Abelian
fusion algebra,
\begin{equation}
\Phi_\lambda \times \Phi_{\lambda'} = \Phi_{\lambda + \lambda'} +
\Phi_{\lambda-\lambda'}
\end{equation}
as long as $\lambda \pm \lambda' \ne 0,p$.   This may be viewed as a
consequence of a simple trigonometric formula. The case where
$\lambda\pm \lambda' = 0$ or $p$ is slightly more subtle. Were these
Luttinger liquid operators, rather than orbifold ones, we would expect
the case $\lambda=\pm\lambda'$ to result in a fusion to the identity, to
$\Phi_{2\lambda}$ and to their descendants. The descendants would then
include $\partial_x\phi_R\partial_x\phi_l$. Due to the orbifold
constraint, this is not a descendant of the identity, and should be
taken into account as a separate fusion product.

The final set of quasiparticles to consider are those in the twisted
sector.   These may be constructed from the local operators
\begin{align}
{\cal V}_5^Q &\sim  e^{i Q \theta_4(x)}  = e^{i  Q
(\bar\phi_{R\rho}-\bar\phi_{L\rho})} {\cal O}_Q^{\sigma_1} \label{o5}\\
{\cal V}_6^Q &\sim  e^{i Q \theta_4(x)} {\cal V}_1  = e^{i  Q
(\bar\phi_{R\rho}-\bar\phi_{L\rho})} {\cal O}_Q^{\sigma_2}
\label{o6}
\end{align}
where $Q$ is an {\it odd} integer identified with (\ref{Q=1},\ref{Q=3})
and the neutral quasiparticle operator ${\cal V}_1$ is given in
(\ref{calv1}).
For $p=1$, ${\cal O}_Q^{\sigma_1}$ has a simple representation in the
unrotated basis: $e^{i Q (\phi_{\sigma R}-\phi_{\sigma L})/4}$.   These
are associated with ${\cal
O}_Q^{\sigma_1}=(\sigma_R^+\sigma_L^-,\tau_R^-\tau_L^+,\tau_R^+
\tau_L^-,\sigma_R^-\sigma_L^+)$ for $Q = (1, 3, 5 , 7)$ mod $8$.   Since
these can be combined with the neutral quasiparticle ${\cal V}_1$ in
(\ref{calv1}) (which in the unrotated basis involves $e^{\pm i
(\phi_{\sigma R}-\phi_{\sigma L})}$) and the trivial quasiparticle
${\cal V}_0$ in (\ref{calv0}), we have ${\cal O}_Q^{\sigma_2} \sim {\cal
O}^{\sigma_1}_{Q\pm 4}$.  Thus, quasiparticles tunneling operators with
$Q=1$ mod 4 will be associated with $\sigma_R^+\sigma_L^-$ or $\tau_R^+
\tau_L^-$, while operators with $Q=3$ mod $4$ will be associated with
$\sigma_R^-\sigma_L^+$ or $\tau_R^- \tau_L^+$.

For $p>1$, there is no longer a simple bosonized representation for the
twist operators.   Nonetheless, since the twist operators in the
orbifold theory retain their identity independent of the orbifold radius
(or $p$), we expect the above identification to remain valid.   The only
complication is that operators differing by ${\cal V}_1$ can no longer
be distinguished.  Thus,
\begin{equation}
{\cal O}_Q^{\sigma_1} = \left\{\begin{array}{ll}
a \ \sigma_R^+\sigma_L^- + b\  \tau_R^+ \tau_L^- & Q = 1\  {\rm mod}\  4
\\
a\  \sigma_R^-\sigma_L^+ + b \ \tau_R^- \tau_L^+ & Q = 3\  {\rm mod}\  4
\end{array}\right.
\label{osigma1}
\end{equation}
where $a$ and $b$ are numerical coefficients.    We note that
$\sigma^\pm$ and $\tau^\pm$ are related by the neutral quasiparticle
$\Theta$: $\tau^\pm = \Theta \sigma^\pm$.   It follows that
${\cal O}_Q^{\sigma_2} = {\cal O}_Q^{\sigma_1}{\cal O}_{\Theta}$ has the
same form as (\ref{osigma1}) with $\sigma$ and $\tau$ interchanged.

\subsection{Bulk Quasiparticle Structure}\label{sec:VI.B}

We now consider the structure of the quasiparticle excitations from the
point of view of the bulk.   In the coupled wire model, bulk
quasiparticles are described by kinks in the fields $\bar\theta_\rho$
defined on the links between wires, as well as corresponding excitations
of the neutral sector.

The structure of the bulk quasiparticle excitations can be characterized
by specifying the set of distinct quasiparticle sectors $a$, along with
data including their quantum dimension, their topological spin, as well
as their braiding statistics.   This data can be summarized by the
topological ${\cal S}$ and ${\cal T}$ matrices\cite{bais2009,
kitaev2006}.   The ${\cal T}$ matrix characterizes the effect of a
$2\pi$ rotation, and is given by
\begin{equation}
{\cal T}_{ab} =  \delta_{ab} e^{2\pi i( h_a - c/24)}
\label{tmatrix}
\end{equation}
where $h_a$ is the dimension of the quasiparticle operator, which we
determined in the previous section using the edge state theory.  In our
theory the central charge is $c=2$, including both the charge and
neutral sectors.

The matrix ${\cal S}$ is a symmetric matrix, where the element ${\cal
S}_{ab}$  characterizes quasiparticles $a$ and $b$ with linked
world lines.   It includes information about both the quantum dimensions
$d_a$, which characterize the multiplicity of fusion channels of the
quasiparticles, as well as the monodromy matrix ${\cal M}_{ab}$, which
characterizes the interference between quasiparticles, and is directly
measured by interferometry using double point contacts.     They are
related by
\begin{equation}
{\cal S}_{ab} =  {\cal M}_{ab}   d_a d_b/{\cal D}
\label{smd}
\end{equation}
where the total quantum dimension satisfies ${\cal D}^2 = \sum_a d_a^2$.
Further properties of the quasiparticles, such as the fusion
coefficients, can be constructed from the $S$ matrix.

We will now show that $d_a$ and ${\cal M}_{ab}$ can be determined in our
theory by considering a long cylinder, which can be modeled as a single
wire with left and right moving chiral modes coupled by electron
tunneling.

\subsubsection{Quasiparticles and Kinks}\label{sec:VI.B.1}

We consider a cylinder modeled by a single wire with right and left
movers locked by the electron tunneling term, which as in
(\ref{ht1thetarhophisigma}) takes the form,
\begin{equation}
V = -4t( \cos 2 l \bar\theta_\rho \cos p \bar\theta_\sigma - s_p \sin 2l
\bar\theta_\rho \cos p \bar\varphi_\sigma).
\label{hcouple}
\end{equation}
This potential has several minima, which correspond to different
topological sectors for the ends of the cylinder.    The minima occur at
\begin{equation}
\bar\theta_\rho = \theta^*_{\rho,Q} = Q \frac{\pi}{4l}
\end{equation}
for integer $Q$.    When $Q$ is an even integer, $\bar\theta_\sigma$ is
pinned at
\begin{equation}
\bar\theta_\sigma = \theta^*_{\sigma,m} = m \frac{\pi}{p}.
\end{equation}
$m$ is even (odd) for $Q=0$ mod $4$ ($Q=2$ mod 4).  When $Q$ is an odd
integer, $\bar\varphi_\sigma$ is pinned at
\begin{equation}
\bar\varphi_\sigma = \varphi^*_{\sigma,n } = n \frac{\pi}{p}.
\end{equation}
where $n$ is even (odd) for $Q-p =0$ mod $4$ ($Q-p = 2$ mod $4$).

Let us suppose that the starting state of the cylinder, in the trivial
state at the end is given by
\begin{equation}
|1\bar 1\rangle = |\theta^*_{\rho,0}\rangle \otimes
|\theta^*_{\sigma,0}\rangle.
\end{equation}
Creating a quasiparticle-antiquasiparticle pair out of the vacuum then
amounts to making kinks in $\bar\theta_\rho$ and $\bar\theta_\sigma$ (or
$\bar\varphi_\sigma$).  Between $a$ and $\bar a$ the cosine potential is
pinned at a different value.    For a quasiparticle with charge $Q
e/(2l)$ the charge sector between the two quasiparticles will be
\begin{equation}
|\theta_{\rho,Q}^*\rangle
\end{equation}
Depending on $Q$ there are allowed minima for $\bar\theta_\sigma$ or
$\bar\varphi_\sigma$, which correspond to different topological sectors
in the neutral sector.   These can be identified with primary fields of
the orbifold theory.

First, consider sectors without kinks.
\begin{align}
&|(1 \bar 1)\rangle =  |\theta^*_{\sigma,0} \rangle, \nonumber\\
 &|(\Theta\bar\Theta)\rangle = |\theta^*_{\sigma,0} \rangle'.
 \label{1bar1}
\end{align}
The difference between $|\theta^*_{\sigma,0} \rangle$ and
$|\theta^*_{\sigma,0} \rangle'$ is that $\Theta$ involves the 
operator $\partial_x\phi$, which is odd under the orbifold symmetry
$\phi \rightarrow -\phi$.   Though there is no kink in
$\bar\theta_\sigma$ this operator can be ``seen" by the twist operators,
which take $\phi$ to $-\phi$.

For the $a= \Psi_\pm$ we have
\begin{align}
|(\Psi_+ \bar \Psi_+)\rangle &= |\theta^*_{\sigma,n=p} \rangle \\
|(\Psi_- \bar \Psi_-)\rangle &= |\theta^*_{\sigma,n=p} \rangle'
\end{align}
Note that $\theta^*_{\sigma,p} = \theta^*_{\sigma,-p}$ mod $2\pi$.
Again, the difference between $\Psi_+$ and $\Psi_-$ is the symmetry
under the orbifold symmetry.   $\Psi_{+(-)}$ is a superposition of
$+\pi$ and $-\pi$ kinks with relative phase $+i (-i)$.

For the $\Phi_k$ operators,
\begin{equation}
|(\Phi_k\bar \Phi_k) \rangle = \frac{1}{\sqrt{2}} (|\theta^*_{\sigma,k}
\rangle +|\theta^*_{\sigma,-k} \rangle)\\
\end{equation}
Here, note that because the state must be symmetric under
$\phi\rightarrow -\phi$, the state is a superposition of two different
kinks, which are physically distinct from one another. Were it not for
the orbifold constraint, a local measurement would have been able to
measure the local value of $\theta^*$, distinguish between the two
components of the superposition, and thus lead to its decoherence.
However, due to the constraints that the orbifold theory imposes on the
allowed operators, such an operator does not exist.   This superposition
indicates a non trivial quantum dimension, as detailed below.

Finally, for the twist operators $\bar\varphi_\sigma$ will be pinned at
one of the $p$ minima of $\pm s_p \cos p\varphi_\sigma$.   One of these
minima will be at either $\varphi_\sigma = 0$ or $\varphi_\sigma =\pi$
(depending on $s_p$ and $Q$).   The other $p-1$ minima  will be a
multiple of $2\pi/p$ away.    Since $\Delta\theta_\sigma \equiv
\theta_\sigma(L) - \theta_\sigma(0) = 0$ and $[\Delta\theta_\sigma,
\varphi_\sigma] = 2\pi i/p$, the states can not be simultaneously
specified by $\varphi_\sigma$ and $\theta_\sigma$.  We can write the
 quasiparticle-antiquasiparticle pair that comes from the identity in
 two different bases:
\begin{align}
|(\sigma^\pm\bar\sigma^\pm) \rangle &=  |\theta^*_{\sigma,0};\ .\
;\theta^*_{\sigma,0}\rangle\\
&= \frac{1}{\sqrt{p}}\sum_{m=-\frac{p-1}{2}}^{\frac{p-1}{2}} |\ .\
;\varphi^*_{\sigma,m s_p};\ .\ \rangle \\
\end{align}
$|( \tau^\pm\bar\tau^\pm) \rangle$ can be expressed similarly, and as in
(\ref{1bar1}) is  distinguished  by how it transforms under the orbifold
symmetry.

\begin{figure}
\includegraphics[width=3in]{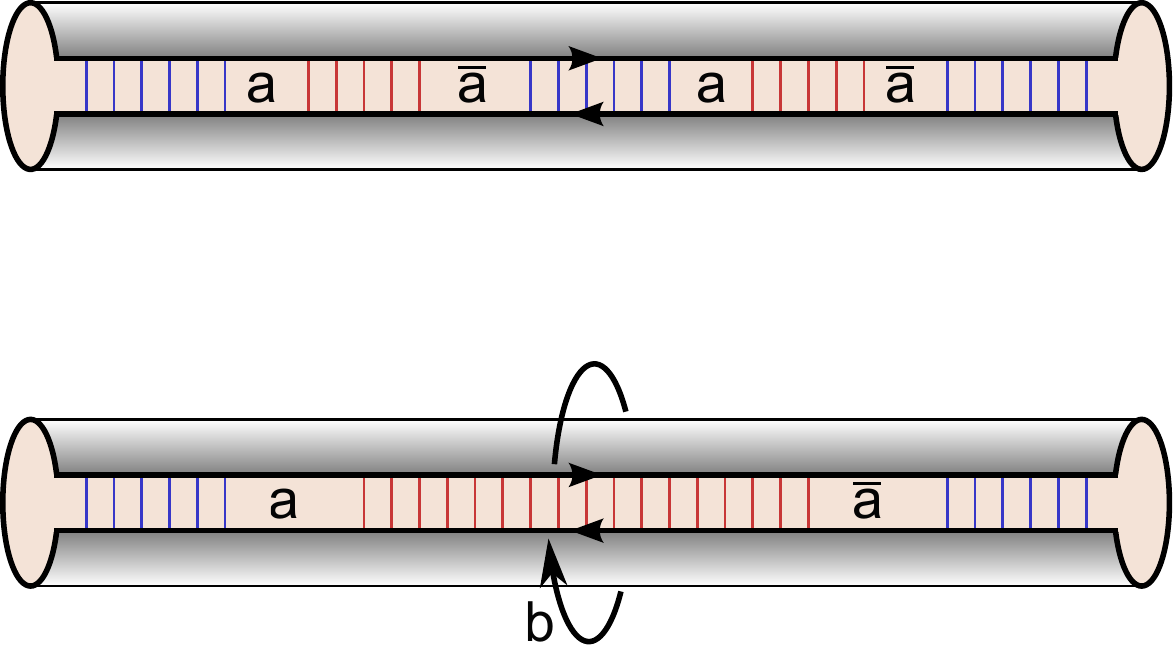}
\caption{A long cylinder modeled by a single wire with right and left
moving modes coupled by electron tunneling as in (\ref{hcouple}).   Bulk
quasiparticles are described by kinks in $\bar\theta_\rho$,
$\bar\theta_\sigma$ and $\bar\varphi_\sigma$.  In (a) two pairs of
quasiparticles ($a\bar a$) are created and recombined in the opposite
order, allowing determination of the quantum dimension $d_a$.   In (b) a
quasiparticle pair ($a\bar a$) is created and a second pair ($b\bar b$)
goes around the cylinder, allowing the determination of the monodromy
matrix $M_{ab}$.}
\label{cylinderfig}
\end{figure}

\subsubsection{Quantum Dimensions}\label{sec:VI.B.2}

The quantum dimension $d_a$ may be determined by the following
construction.    Create two sets of quasiparticle- anti quasiparticle
pairs $a, \bar a$, and then bring the middle pair together followed by
bringing the outer pair together.    If $a$ and $\bar a$ have multiple
fusion channels, then they need not fuse to the identity.   The
probability amplitude that the system returns to the ground state will
be given by $1/d_a$.   Equivalently, if we define $|(\bar a a)(\bar a
a)\rangle$ as the state where two separated pairs are created from the
identity, and $|(\bar a (a \bar a) a)\rangle$ as the state where the
second pair $(a\bar a)$ is created between the first pair $(\bar a a)$,
then
\begin{equation}
d_a = |\langle( a (\bar a a) \bar a)|(a \bar a)(a \bar a)\rangle |^{-1}.
\end{equation}

We now consider the construction where we create two $a\bar a$ pairs as
in Fig. \ref{cylinderfig}(a) and annihilate them in the opposite order.
For $a=1, \psi^\pm, \Theta$, each $\bar a a$ pair defines a pure state,
so the amplitude to get back to the ground state is $1$, and $d_a=1$.

For $a=\phi_k$, we can write the state with two pairs as (suppressing
the $\sigma$ subscript for brevity)
\begin{eqnarray}
|(\phi_k \bar \phi_k)(\phi_k \bar \phi_k)\rangle &=&
 \frac{1}{2}(|\theta^*_k;\theta^*_0;\theta^*_k\rangle +
 |\theta^*_{-k};\theta^*_0;\theta^*_k;\rangle \\
 &+&
 |\theta^*_k;\theta^*_0;\theta^*_{-k}\rangle+|\theta^*_{-k};\theta^*_0;\theta^*_{-k};\rangle)\nonumber
\end{eqnarray}
On the other hand, the state with the quasiparticles paired in the
opposite order will be
\begin{eqnarray}
|(\phi_k (\bar \phi_k\phi_k )\bar \phi_k)\rangle &=
 &\frac{1}{2}(|\theta^*_k;\theta^*_0;\theta^*_k\rangle +
 |\theta^*_{k};\theta^*_{2k}\theta^*_k;\rangle \\
 &+&
 |\theta^*_{-k};\theta^*_0;\theta^*_{-k}\rangle+|\theta^*_{-k};\theta^*_{-2k};\theta^*_{-k};\rangle)\nonumber
\end{eqnarray}
It can be seen that the overlap is $1/2$, so
\begin{equation}
d_{\phi_k}=2.
\end{equation}
Finally, for the twist quasiparticles, $\varphi_\sigma$ is pinned within
each pair, but $\theta_\sigma$ is pinned between the pairs.   Since
these operators do not commute with one another, they can not be
simultaneously specified.   The state can be expressed either in a basis
of eigenstates of $\theta_\rho$ or of the two $\varphi_\sigma$'s.   For
the first ordering we have,
\begin{eqnarray}
|(\sigma^+\bar\sigma^+)(\sigma^+\bar\sigma^+)\rangle = &|&\cdot
;\theta^*_0 ; \cdot\rangle  \nonumber\\
= \frac{1}{p}\sum_{mn} &|&\varphi_{m s_p}^*;\cdot ;\varphi_{n
s_p}^*\rangle
\end{eqnarray}
For the other order, the second pair of quasiparticles with
$\theta_\sigma$ pinned sits between $\varphi_\sigma$ eigenstates with
$\Delta\varphi=0$, leading to an equal amplitude superposition of the
$\theta^*_m$ states,
\begin{eqnarray}
|(\sigma^+(\bar\sigma^+\sigma^+)\bar\sigma^+)\rangle &=&
\frac{1}{\sqrt{p}}\sum_{m} |\varphi_{m s_p}^*;\cdot ;\varphi_{m
s_p}^*\rangle    \nonumber  \\
&=& \frac{1}{\sqrt{p}}\sum_{m} |\cdot;\theta_m^*;\cdot\rangle
\end{eqnarray}

The overlap is $1/\sqrt{p}$, and a similar result is obtained for the
other twist quasiparticles.   We thus conclude
\begin{equation}
d_{\sigma^\pm} = d_{\tau^\pm} = \sqrt{p}.
\end{equation}

This quantum dimension is reminiscent of that of the $Z_p$ parafermions
found on counter-propagating $\nu=\pm 1/p$ edge modes gapped in an
alternating way by superconductors and normal backscattering. In that
case, however, the presence of Cooper-pairs makes the $Z_p$ parafermion
occur always with a $Z_2$ Majorana zero mode.

\subsubsection{Monodromy Matrix}\label{sec:VI.B.3}

To determine the monodromy matrix $M_{ab}$ we create a
quasiparticle-antiquasiparticle pair $a,\bar a$.   Then, between $a$ and
$\bar a$ we create a pair $b, \bar b$ and take $b$ around the cylinder.
$M_{ab}$ compares the probability amplitude for $b$ and $\bar b$ to fuse
to the identity when taken around $a$ to that with $a=1$.
\begin{equation}
{\cal M}_{ab} = \frac{\langle (a\bar a)| {\cal O}_b | (a\bar
a)\rangle}{\langle (1\bar 1)|{\cal O}_b|(1\bar 1)\rangle}
\end{equation}

Obviously for $b=1$, $M_{a1}=1$.   Consider next terms with
$b=\psi^\pm$.  The operator $
{\cal O}_{\psi^\pm} = (\Psi^\pm_L)^\dagger \Psi^\pm_R$
can be evaluated using the analysis that led to (\ref{psiproduct}) to be
 \begin{equation}
{\cal O}_{\psi^\pm} = \cos p\theta_\sigma + i s_p  \sin p\varphi_\sigma
\end{equation}
We then simply evaluate the expectation value of this operator at the
appropriate minimum, characterized by $\theta^*_m$ or $\varphi^*_{n
s_p}$.    For the non twist operators we find
\begin{eqnarray}
{\cal M}_{1\psi^\pm } &=& M_{\psi^\pm \Theta}=1   \nonumber \\
{\cal M}_{ \psi^+\psi^\pm}&=& M_{\psi^\pm \psi^-} = -1\\
{\cal M}_{\phi_k\psi^\pm } &=& (-1)^k   \nonumber
\end{eqnarray}
For the twist operator we find,
\begin{eqnarray}
{\cal M}_{ \sigma^+\psi^\pm} = M_{\tau^+\psi^\pm } = \pm i s_p
\nonumber \\
{\cal M}_{\sigma^-\psi^\pm } = M_{\tau^-\psi^\pm } = \mp i s_p
\end{eqnarray}

For $b = \Theta$, ${\cal O}_{\Theta} = \partial_x\phi_{\sigma,R}
\partial_x\phi_{\sigma,L}$.   It follows that for the non twist
operators (where $\theta_\sigma=\phi_{\sigma,R}-\phi_{\sigma,L}$ is
pinned) $\partial_x\phi_{\sigma,R} = \partial_x\phi_{\sigma,L}$, so
$\Theta$ is unaffected when taken around the cylinder.   In contrast,
for the twist operators (where
$\varphi_\sigma=\phi_{\sigma,R}+\phi_{\sigma,L}$ is pinned)
$\partial_x\phi_{\sigma,R} = -\partial_x\phi_{\sigma,L}$, so $\Theta$
changes sign when taken around the cylinder.   We thus conclude that
\begin{eqnarray}
&{\cal M}_{ 1\Theta} = {\cal M}_{\psi^\pm\Theta }={\cal
M}_{\Theta\Theta} = {\cal M}_{\phi_k\Theta } = 1,  \nonumber \\
&{\cal M}_{\sigma^\pm\Theta } = {\cal M}_{\tau^\pm\Theta } = -1.
\end{eqnarray}

For $b=\phi_k$, ${\cal O}_{\phi_k} = \cos k \theta_\sigma$.  For the non
twist operators,
\begin{eqnarray}
{\cal M}_{1\phi_k } &=& {\cal M}_{ \Theta\phi_k} = 1  \nonumber \\
{\cal M}_{\psi^\pm\phi_k } &=& (-1)^k \\
{\cal M}_{\phi_{k'}\phi_k } &=& \cos \pi k k'/p. \nonumber
\end{eqnarray}
For the twist operators, since $\varphi_\sigma$ is pinned,
\begin{equation}
{\cal M}_{\sigma^\pm\phi_k } = {\cal M}_{\tau^\pm\phi_k } = 0.
\end{equation}

It remains to determine the ``twist-twist" components of ${\cal
M}_{ab}$.   Since our bosonization approach does not provide an explicit
formula for the twist operators when $p>1$ we do not have a simple
calculation for these terms.   Nonetheless, the ${\cal S}$ matrix for
the orbifold theory is well known in the conformal field theory
literature\cite{dijkgraaf1989}.   Here we will use that result and point
out a minor subtlety associated with the ${\cal S}$ matrix derived in
Ref. \onlinecite{dijkgraaf1989}.

The ${\cal S}$ matrix is shown  in Table \ref{smatrixtab}.  It can be
seen that for all of the entries that involve a non-twist field, it
agrees with our calculation of $d_a$ and ${\cal M}_{ab}$ using
(\ref{smd}).    It is equivalent to the ${\cal S}$ matrix quoted in Ref.
\onlinecite{dijkgraaf1989} (which we will call ${\cal S}_{DVVV}$),
except for the presence of the factors $s_p$, which are $-1$ for $p=1$
mod $4$ and $+1$ for $p=3$ mod $4$.   This changes the sign of the
imaginary part of ${\cal S}$ when $p=1$ mod 4, so that
\begin{equation}
{\cal S}_{DVVV} = \left\{\begin{array}{ll} {\cal S}^* & p=1\ {\rm mod} \
4 \\
{\cal S} & p = 3 \ {\rm mod}\ 4. \end{array}\right.
\end{equation}
The ${\cal S}$ matrix defined in Table \ref{smatrixtab}, along with the
${\cal T}$ matrix defined in (\ref{tmatrix}) satisfy the general
constraint of modular invariance\cite{byb, kitaev2006} $({\cal S} {\cal
T})^3 = {\cal C}$, where ${\cal C}$ is the charge conjugation matrix,
which takes particles to antiparticles.   When $p=1$ mod 4, this
relation is {\it not} satisfied by ${\cal S}_{DVVV}$, but rather by
${\cal S}_{DVVV}^*$.  In this case, $({\cal S}_{DVVV} {\cal T})^3 = 1$,
which can be seen by noting that ${\cal S}^* = {\cal C} {\cal S}$.

The minor modification ${\cal S} \rightarrow {\cal S}^*$ could be viewed
as book keeping, since in general ${\cal S}^*$ describes a time reversed
system that has edge states that propagate in the opposite direction but
is otherwise the same.   However, time reversal also takes ${\cal T}
\rightarrow {\cal T}^*$, so to be consistent ${\cal T}_{DVVV}$ (which is
not displayed in Ref. \onlinecite{dijkgraaf1989}) would have to be
${\cal T}_{DVVV} = {\cal T}^*$ for $p=1$ mod $4$.     Since in principle
both ${\cal S}$ and ${\cal T}$ are measurable, this distinction has
physical consequence.

\begin{table}
  \centering
\begin{ruledtabular}
\begin{tabular}{c|ccc|c|cc}
            & $1$ & $\Theta$ & $\psi^s$ & $\phi^k$ & $\sigma^s$ &
            $\tau^s$ \\
 \hline
$1$           & $1$ & $1$      & $1$      & $2$      & $\sqrt{p}$
&$\sqrt{p}$    \\
$\Theta$      & $1$ &  $1$     &  $1$     &  $2$     &   $-\sqrt{p}$
&  $-\sqrt{p}$  \\
$\psi^{s'}$   & $1$ & $1$     & $-1$      & $2(-1)^k$ &   $i s s'
s_p\sqrt{p}$   &  $i s s' s_p \sqrt{p}$ \\
\hline
$\phi^{k'}$   & $2$ & $2$& $2(-1)^{k'}$    & $4\cos\frac{\pi k k'}{p}$
&  $0$  & $0$  \\
\hline
$\sigma^{s'}$ & $\sqrt{p}$ & $-\sqrt{p}$& $i s s' s_p \sqrt{p}$  & $ 0$
&  $\sqrt{p}e^{i\frac{\pi}{4} s s' s_p }$  &  $-
\sqrt{p}e^{i\frac{\pi}{4} s s' s_p}$ \\
$\tau^{s'}$   &  $\sqrt{p}$&$-\sqrt{p}$ & $i s s' s_p\sqrt{p}$ &  $0$ &
$- \sqrt{p}e^{i \frac{\pi}{4}s s' s_p}$  &  $\sqrt{p}e^{i \frac{\pi}{4}s
s' s_p}$ \\
\end{tabular}
\end{ruledtabular}
\caption{S-matrix.  $s$ and $s'$ are $\pm 1$, and $k = 1, ..., p-1$.}
\label{smatrixtab}
\end{table}

\section{Discussion}\label{sec:VII}

In this concluding section we will begin by discussing two special cases
of the orbifold theory that can be understood using simpler methods.
We will then conclude with some comments on extensions and open
problems.

\subsection{Special Cases}\label{sec:VII.A}

It is well known\cite{dijkgraaf1989} that the orbifold theory at $p=1$
corresponds to an Abelian $U(1)_8$ theory, and that the theory at $p=3$
corresponds to the $Z_4$ parafermion theory, which can be represented as
a $SU(4)_4/U(1)$ coset.    In this section we show how these facts can
be understood in our coupled wire construction.   These
interconnections, which were also noted in Ref.
\onlinecite{BarkeshliWen2011}, provide deeper insight into the nature of
the orbifold states.

\subsubsection{p=1:  Abelian States}\label{sec:VII.A.1}

For $p=1$, the solvable point, where the charge $e$ tunneling operators
defined in (\ref{tildepsiml})
are purely chiral corresponds to the point $\tilde g_\sigma = 4$ in
(\ref{4/p}).   From
(\ref{gbsigma})  it can be seen that this corresponds to
$\lambda_\sigma = 0$ in (\ref{hupdown2},\ref{hsigrotate}) and   $u=0$ in
(\ref{hsigunrotate}).   Thus, for $p=1$, the nonlinear term in the
unrotated basis vanishes, so the theory is equivalent to the coupled
wire model for a two component Abelian fractional quantum Hall state
defined at filling factor
\begin{equation}
\nu = 2/l,
\end{equation}
where $l = 2m-1$ for odd integers, $m$, so that
\begin{equation}
l = ..., -3, 1, 5, 7, ...
\end{equation}
This Abelian state can be understood in two ways.   In the unrotated
basis, the analysis is similar to that in Section II.D of Ref.
\onlinecite{teokane2014} and leads to an Abelian theory with a  $2\times
2$ $K$ matrix.   This Abelian theory can be interpreted as a constrained
product of an Abelian charge sector with $K=4l$ and a Abelian neutral
sector with $K=8$.    In the rotated basis, the charge sector is
unchanged, but the $K=8$ theory is replaced by the orbifold theory at
$p=1$.    Here we present the translation between those two points of
view by first describing the Abelian theory and then comparing it to the
orbifold theory.

The  elementary local operators for the edge states are the chiral
charge $e$ and charge $4e$ operators, which are determined by
(\ref{psiml}) and (\ref{PSIML}) with $L = 4l$ and $M=l$.   They
 may be written in the canonical form
\begin{eqnarray}
\psi_{e,R/L}  &=&  e^{i \sum_b K_{1b} \bar\phi_{b,R/L}} \\
\Psi_{4e,R/L} &=& e^{i \sum_b K_{2b} \bar\phi_{b,R/L}},
\end{eqnarray}
where the $2\times 2$ $K$ matrix will be determined by the relation
\begin{equation}
\left[\partial_x\bar\phi_{a,A}(x),\bar\phi_{b,A'}(x')\right] = 2\pi i
\tau^z_{AA'} K^{-1}_{ab} \delta(x-x').
\label{pbcomm}
\end{equation}
The fields in the exponents have the form
\begin{equation}
\sum_{b=1}^2 K_{ab} \bar\phi_{A,b} = \sum_{k=1}^4 M^A_{ak} \Phi_k,
\end{equation}
where $A=R/L$ and  $\Phi_k = (\phi_{1R},\phi_{1L},\varphi_4,\theta_4)$
are the elementary fields defined in (\ref{phivec}).
Using (\ref{psiml}) and (\ref{PSIML}), $M^{R/L}_{ak}$ can be determined
to be
\begin{equation}
\left(\begin{array}{cc} M^R_{ak} \\ \\ M^L_{ak}\end{array} \\ \right) =
\left(\begin{array}{cccc}
(3+l)/4 &  (1-l)/4 &0& l \\
l & -l &1& 4l \\
(1-l)/4& (3+l)/4 & 0 & -l\\
-l & l & 1& -4l
\end{array}\right),
\end{equation}
where again we note that $m=(1+l)/2$ is an odd integer.
Using the commutation relations obeyed by $\Phi_k$, Eq. \ref{pbcomm}
follows with
\begin{equation}
K = \left(\begin{array}{cc}  (l+1)/2 &  2 l \\ 2 l& 8 l
\end{array}\right).
\label{Mmatrixp1}
\end{equation}
In terms of these fields the chiral charge density is given by $\rho_R =
e\sum_a t_a \partial_x \bar\phi_{a,R}/(2\pi)$, with the charge vector
${\bf t} = (1,4)^T$.   It can be checked that the filling factor
satisfies $\nu = {\bf t}^T\cdot K^{-1} \cdot {\bf t}$. Interestingly,
the electronic contribution to the filling factor, which is ${\bf
t_e}^TK^{-1} {\bf t}$ with ${\bf t_e}=(1,0)^T$, vanishes. This is in
accordance with our assumption that  the single wires are at their
critical point, at which the density of single electrons vanishes.
Despite their vanishing density, the presence of electrons as local
degrees of freedom makes the state different from that of bosonic
$4e$-clusters.

Quasiparticle operators are given by
$\exp(i\sum_{a} n_a \bar\phi_a)$, where $n_a$ is an integer valued
vector.   The number of independent quasiparticle sectors (and hence the
ground state degeneracy on a torus) is determined by ${\rm Det}(K) =
4l$. The smallest charge quasi-particle has a charge $1/2l$ and there
are two topologically different quasi-particles at each charge $j/2l$,
with $j=1,...,2l$. The two identical-charge quasi-particles may be
transformed to one another by fusion with a neutral quasi-particle for
which $n=(1,0)^T$. This quasi-particle has bosonic self statistics, but
accumulates a phase $\pi j$ when encircling a quasi-particle of charge
$j/2l$.

To make contact with the representation of the system in terms of the
$SU(2)$ fermions, it is instructive to transform from the fields
$\bar\phi$ to the bosonic fields $\phi_{\uparrow, R(L)},
\phi_{\downarrow, R(L)}$ that describe the $SU(2)$ fermions. We carry out
the following transformation (focusing on the right-moving edge)
\begin{align}
{\bar\phi}_{1,R}&=\phi_{\uparrow R}-\phi_{\downarrow R} \nonumber \\
{\bar\phi}_{2,R}&=\frac{1}{2}\phi_{\downarrow R}.
\label{peq1fermions}
\end{align}

Under this transformation the K-matrix becomes
\begin{equation}
{\tilde K}=\left (\begin{matrix} (l+1)/2 & (l-1)/2 \\ (l-1)/2 & (l+1)/2
\end{matrix}\right),
\end{equation}
while the charge vector becomes ${\tilde t}=(1,1)$. This K-matrix and
charge vector correspond to a fractional quantum Hall state of spinful
composite fermions at filling factor $2$. In this state the composite fermions fill one up-spin and one
down-spin Landau levels.  The transformation to the composite fermions
is carried out by the attachment of the even number $m-1 = (l-1)/2$ flux
quanta to each fermion.  The new K-matrix and charge vector should,
however, be used with caution. The transformation (\ref{peq1fermions})
has a determinant of $1/2$.   As such, the ground state degeneracy is
$4\det {\tilde K}$, and the charge of the lowest quasi-particle charge
is $e^*/e=\frac{1}{2}{\min_{ l}l^T{\tilde K}^{-1}{\tilde t}}$, with $l$
being integer valued vectors.

Following through this transformation, we can describe all
quasiparticles in terms of the $SU(2)$ fermions. This is easiest to
exemplify on the $l=1$ case, corresponding to $\nu=2$. Quasiparticles
must be local with respect to the electron, i.e., must accumulate an
integer number of $2\pi$ phase when encircling the electron. The
electron creation operator, describing a local degree of freedom of
charge one, must be composed of an even number of spin-down fermion
operators and an odd number of spin-up fermion operators. The exclusion
of an odd number of spin-down fermions from the physical Hilbert space,
which is the crucial difference between the problem we deal with and
``conventional" chiral fermions, allows for an operator of half a spin
down fermion to be local with respect to the electron, and hence be a quasi-particle (as indeed shown by the second line
of Eq. (\ref{peq1fermions})). Thus, to be local with respect to the
electron an excitation needs to have any number of half-integer spin
down fermions, and an integer number of spin-up fermions. There are
eight topologically distinct excitations of this type, with the number
of spin-down fermions being $0,1/2,1,3/2$ and the number of spin-up ones
being $0,1$. The neutral quasi-particle that is topologically distinct
from the vacuum is a charge-zero spin-one quasi-particle, namely a
spin-up fermion with a spin-down hole. In the $SU(2)$ fermions language in
the unrotated basis, the topologically non-trivial nature of this
excitation is a consequence of its violation of the constraint that
forces an even number of spin-down fermions on each wire.

To make contact with the orbifold description it is useful to recast the
$K$ matrix in the charge-neutral basis.  To this end, we express the
elementary operators in terms of
\begin{eqnarray}
\bar\phi_{\rho,R/L} &=& \phi_{1,R/L}/4+\phi_{2,R/L}\\
\bar\phi_{\sigma,R/L} &=& \phi_{1,R/L}/4,
\end{eqnarray}
In the language of the $SU(2)$ fermions, the smallest local object in the
charge sector is composed of two pairs of fermions in a singlet state
(total charge $4$), described by
 $e^{i8\bar\phi_{\rho,R/L}}$. The smallest local object in the spin
 sector is composed of two charge-zero spin-1 excitations,  described by
 $e^{i8\bar\phi_{\sigma,R/L}}$. In each case, locality requires an
 operator to have an even number of spin down creation operators. An
 electron is a product of a quarter of the local charge and spin
 excitations.

\begin{figure}
\includegraphics[width=2in]{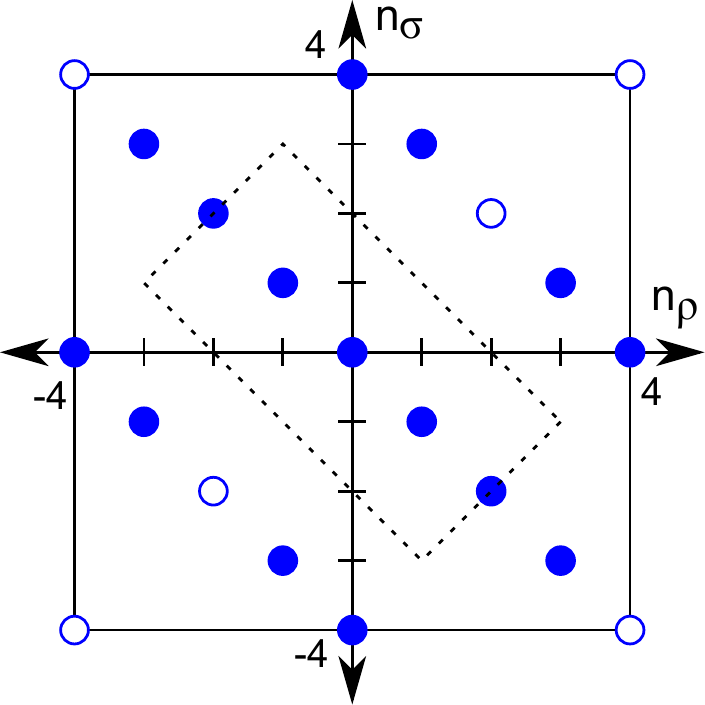}
\caption{Quasiparticles of the $\nu=2$ state with $p=l=1$ in the
unrotated basis.   The quasiparticles can be viewed as a combination
charge and neutral operators, each described by Abelian $K=8$ theories
with $n_{\rho,\sigma}$ defined modulo $8$, subject to the constraint
that $n_\rho+n_\sigma$ is a multiple of $4$.    The local operators
built from the charge $e$ and charge $4e$ operators in (\ref{psiandPsi})
are indicated by the open circles.   This leaves $4$ independent
quasiparticle types, as indicated by the dashed rectangle.}
\label{qpabelian}
\end{figure}

In the charge-spin $(\rho\sigma)$ basis
\begin{equation}
K^{(\rho\sigma)} = \left(\begin{array}{cc}  8l & 0 \\ 0 & 8
\end{array}\right)
\end{equation}
In this representation, the charge sector is described by an Abelian
theory with $K_\rho = 8l$, while the neutral sector is the Abelian
theory with $K_\sigma=8$.   Importantly, however, the charge and neutral
sectors are not independent.   The above transformation involves a
factor of $1/4$, so that operators of the form
\begin{equation}
O_{n_\rho,n_\sigma} = e^{i n_\rho \bar\phi_\rho + n_\sigma
\bar\phi_\sigma}
\label{oprhosigma}
\end{equation}
correspond to physical charge $n_\rho e/2l$ quasiparticle operators only
if $n_\rho + n_\sigma$ is a multiple of $4$.   Moreover, the local
charge $e$ and $4e$ operators are given by
\begin{eqnarray}
\psi_{e,R/L} &\sim & e^{i 2l \bar\phi_{\rho,R/L} + 2
\bar\phi_{\sigma,R/L}} \nonumber\\
\Psi_{4e,R/L} &\sim & e^{i 8l \bar\phi_{\rho,R/L}} \label{psiandPsi}
\end{eqnarray}
It follows that the distinct quasiparticle types can be identified with
$-l \le n_\rho < l$, and $n_\sigma = -n_\rho\  {\rm mod }\ 8$ or
$n_\sigma = 4-n_\rho \  {\rm mod } \ 8$.  This gives a total of $4l$
distinct quasiparticle types in agreement with the above count.   The
lattice of quasiparticle sectors for the special case $l=1$ ($\nu=2$) is
indicated in Fig. \ref{qpabelian}.

In general, quasiparticle operators are characterized by their scaling
dimension - or equivalently topological spin.   When decomposed into
charge and neutral components, we write
\begin{equation}
\Delta = \Delta^\rho + \Delta^\sigma
\end{equation}
with
\begin{equation}
\Delta_\rho = \frac{n_\rho^2}{16l}; \quad\quad \Delta_\sigma =
\frac{n_\sigma^2}{16}.
\end{equation}
In particular, the neutral sector is characterized by the ``$K=8$" (or
$U(1)_8$) theory with 8 independent primary fields indexed by $n_\sigma$
modulo $8$.   The Table \ref{qptab} lists these neutral primary fields
with dimension $\Delta_\sigma$, and identifies them  with the primary
fields of the orbifold theory discussed below.

Finally, we note that bulk quasiparticles can be described as kinks in
the pinned bulk bosonic fields defined on the links between wires.   The
tunneling terms can be written in the form
\begin{eqnarray}
H_{4T} &=& t_4 \cos 8l \bar\theta_{i+1/2} \\
H_{1T} &=& t_1 \cos (2l \bar\theta_{\rho, i+1/2} +2
\bar\theta_{\sigma,i+1/2})
\end{eqnarray}
where $\bar\theta_{\rho/\sigma,i+1/2} = \bar\phi_{R,\rho/\sigma,i+1} -
\bar\phi_{L,\rho/\sigma,i}$.    As a function of
$\bar\theta_{\rho,i+1/2}$ and $\bar\theta_{\sigma, i+1/2}$ this leads to
a periodic potential with minima with the pattern shown in Fig.
\ref{qpabelian}, provided we identify $\theta_\rho = 2\pi n_\rho/(8l)$
and $\theta_\sigma = 2\pi n_\sigma/8$.    The kinks that connect those
minima are then precisely the charge $n_\rho/2l$ quasiparticles.

The $SU(2)$ rotation introduced in Section \ref{sec:IV.B} does not
affect the charge sector, while the neutral sector is affected through a
rotation of the spin-axis by $\pi/2$. After the rotation the various
quasi-particles may be identified with their orbifold counterparts, as
shown in Table \ref{qptab}.

\begin{table}
\centering
\begin{ruledtabular}
\begin{tabular}{c|cccccccc}
$n_\sigma$ & $-3$ & $-2$ & $-1$ & $0$ & $1$ & $2$ & $3$ & $4$ \\
${\cal O}_\sigma$ & $\tau^+$ & $\psi^-$ & $\sigma^-$ & $1$ & $\sigma^+$
& $\psi^+$ & $\tau^-$ & $\Theta$ \\
$\Delta_\sigma$ & $9/16$ & $ 1/4$ & $1/16$ & $0$ & $1/16$ & $1/4$ &
$9/16$ & $1$
\end{tabular}
\end{ruledtabular}
\caption{Primary fields of the neutral sector with dimension
$\Delta_\sigma$ in the unrotated and the rotated basis.   In the
unrotated basis, described by an Abelian $U(1)_8$ theory are indexed by
$n_\sigma$ modulo $8$.   In the rotated basis, the same fields are
identified as primary fields ${\cal O}_\sigma$ of the $p=1$ orbifold
theory.}
\label{qptab}
\end{table}

\subsubsection{$p=3$:  the $k=4$ Read Rezayi State}\label{sec:VII.A.2}

\begin{figure}
\includegraphics[width=3.4in]{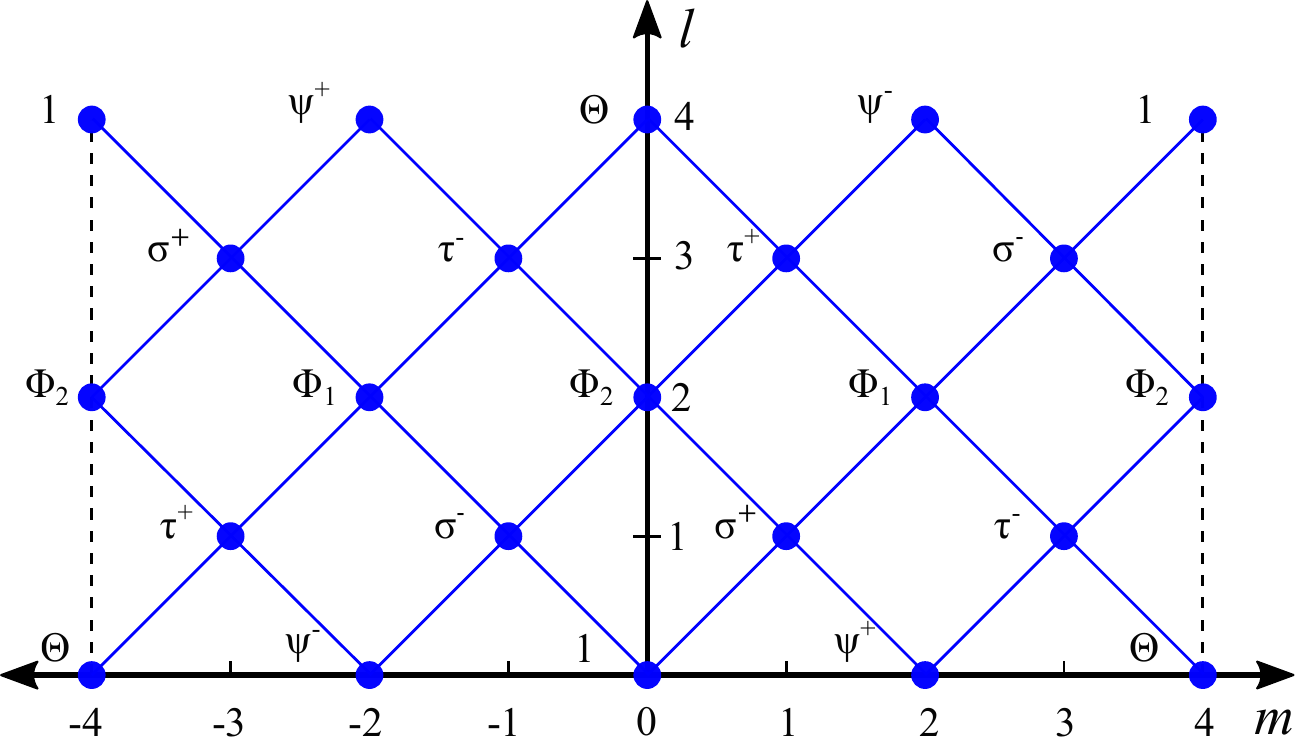}
\caption{Brattelli diagram of the quasi-particles in the $p=3$ state,
which corresponds to the $k=4$ Read-Rezayi state described by the $Z_4$
parafermion theory, $SU(2)_4/U(1)$.   The quasiparticles are indexed by
the integers $m$ and $l$.   $m$, defined modulo $8$ plays a role similar
to $n_\sigma$ above, and is determined by the quasiparticle charge.
$l$ identifies distinct quasiparticle types for a given charge.   The
fusion outcomes of any quasiparticle with $\sigma^+$ can be deduced by
moving one step up or down and to the right.
 }
\label{qpReadRezayi}
\end{figure}

The case of $p=3$ is one of the Read-Rezayi series of non-Abelian states
that are based on electron clustering. The Read-Rezayi series is
composed of the states $\nu=k/(mk+2)$, where $k$ is an integer that
signifies the number of electrons in a cluster and $m$ is an odd integer
(similar states may be constructed for bosons, in which case $m$
is even).  For $k=4$ the
Read-Rezayi series is given by Eq. (\ref{p=3filling}). The
quasi-particles found by identifying the $Z_4$ parafermion CFT of the
Read-Rezayi approach are identical to those found using our approach.
Each quasi-particle is a product of a vertex operator in a free boson
charged mode and an operator that acts on a neutral mode. The latter are
commonly described by the notation $\Phi_m^l$ in which $l=0,...,4$, the
indices $l+m=0 \ ({\rm mod}\ 2)$, and the following identifications
$\Phi_m^l\equiv\Phi_{m+8}^l\equiv \Phi_{m-4}^{4-l}$.  The $Z_4$
parafermion fields, which in our notation are $1$, $\Psi^+, \Theta,
\Psi^-$ are here $\Phi_{2j}^0$, where $j=0,1,2,3$ respectively. The
twist operators  are identified in the following way
$\sigma^\pm=\Phi_{\pm 1}^1$, and $ \tau^\pm=\phi_{\pm 1}^3$.   The two
remaining fields are $\Phi_{\lambda=1}=\Phi_2^2$ and $\Phi_{\lambda=2}
=\Phi_0^2$.   The fusion rules of the various quasi-particles are
described by the Brattelli diagram (see Fig. \ref{qpReadRezayi}) and
reproduce the fusion rules of the $p=3$ orbifold theory.

\subsection{Concluding remarks}\label{sec:VII.B}

In this paper we have introduced a coupled wire model for the $Z_4$
orbifold quantum Hall states, based on a theory of clustering of
electrons into charge $4e$ bosons.    On a single wire we employ a
mapping to the critical point of the four-state clock model, which
exhibits a line of critical points described by the orbifold CFT.
Coupling the wires together to form gapped quantum Hall states leads to
a sequence of quantum Hall states with an Abelian charge sector coupled
to a neutral sector described by the orbifold CFT at a set of discrete
radii parametrized by the odd integer $p$.   For each odd integer $p$
we identify a solvable theory in which the interactions are tuned to
make the electron operators on each wire purely chiral, so that nearest
neighbor tunneling, when relevant, necessarily opens a gap.

As discussed in the previous section, $p=1$ is equivalent to an Abelian
state, while $p=3$ is equivalent to the $Z_4$ parafermion Read-Rezayi
state.   Larger odd values of $p$ define a set of quantum Hall states
that retain the $Z_4$ character, but have additional quasiparticle types
$\Phi_\lambda$.   This sequence of orbifold states characterized by the
odd integer $p$ has a structure reminiscent of the Laughlin sequence at
$\nu=1/m$ for odd integer $m$, which has a similar quasiparticle
structure.   The difference is that the orbifold states feature the
additional twist operators, which lead to a richer non-Abelian
structure.

In this paper we have focused exclusively on the orbifold states defined
for {\it odd} integers $p$.   When $p$ is even, our construction breaks
down because when $l = 2m-p$ is even the electron operator $\psi_{ml}$
in Eq. \ref{psimlboson} necessarily involves the twist operators
$\sigma_R \sigma_L$.   Since the twist operators retain their identity
independent of the orbifold radius, it is not possible to add a forward
scattering interaction that modifies the orbifold radius and makes
$\psi_{ml}$ a purely chiral operator.    Thus, though it is possible for
even $p$ to write an electron tunneling term that involves the twist
operators, we are not able to find a solvable point with an energy gap.
It seems unlikely that such a Hamiltonian would lead to a gapped quantum
Hall state.

This leads to interesting questions for further inquiry.   One question
is what is the nature of the ground state of a clustered coupled wire
model with even $p$, when the electron tunneling operator involves the
twist operators.   If it is not a gapped quantum Hall state, then
perhaps it is an interesting {\it gapless} state.

Secondly, one can ask whether quantum Hall states characterized by
orbifold conformal field theories with even $p$ are possible.   The
answer to this is certainly yes.  As noted in Ref.
\onlinecite{dijkgraaf1989}, the orbifold theory is a well defined
rational CFT when $p$ is even.  This led Barkeshli and
Wen\cite{BarkeshliWen2011} to propose even $p$ orbifold states.
However, the fusion rules for even $p$ have a different form than for
odd $p$, and have a $Z_2 \times Z_2$ structure, rather than a $Z_4$
structure.   It therefore seems likely that the physics of the even $p$
orbifold states is not based on clustering of charge $4e$ bosons, but
rather involves pairs of charge $2e$ bosons.   Analysis of these states
will be left for future work.

A further direction is to ask whether the insights gained from the
orbifold theory can be applied to clustered states with $k\ne 4$.   One
interesting approach is the $\epsilon$ expansion in Ref.
\onlinecite{Sagi2017}, which identified parafermion critical points in
an expansion about $k=4+\epsilon$.

Overall, our work demonstrates the power of the coupled wire approach as
a tool to generate gapped quantum Hall states and study their
topological properties. This tool complements methods employing tools
such as analytic single particle wave functions, Jack Polynomials and
the thin torus limit \cite{RevModPhysHansson}.

\acknowledgments
We thank Erez Berg, Yuval Oreg, Eran Sagi, Andreas Ludwig and Yichen Hu
for helpful discussions.
 This work was supported in part by  grants from the Microsoft
 Corporation and  the US-Israel Binational Science Foundation (AS),  the
 European
Research Council under the European Unions
Seventh Framework Program (FP7/2007-2013) / ERC
Project MUNATOP, the DFG (CRC/Transregio 183, EI
519/7-1), Minerva foundation (AS) and a Simons Investigator grant from
the Simons Foundation (CLK).

\appendix

\section{Relevant Electron Tunneling Operators}\label{appendix:A}

In Section \ref{sec:V.B} we observed that the electron operators $\tilde
\psi_{ml}^\dagger$ defined in (\ref{tildepsiml}) have dimension $\Delta
= \Delta_\rho + \Delta_\sigma$ with $\Delta_\rho = |l|/4$ and
$\Delta_\sigma = p/4$.   Without interactions, the electron tunneling
term will be irrelevant when $2\Delta > 2$.    By introducing a local
interaction term (\ref{hlambdarho}) proportional to $\partial_x
\bar\phi_{i,\rho,R} \partial_x\bar\phi_{i+1,\rho,L}$ we argued that
$\Delta_\rho$ can be made arbitrarily small.    However, the
corresponding operator in the neutral sector, $\partial_x
\bar\phi_{i,\sigma,R} \partial_x\bar\phi_{i+1,\sigma,L}$, which has the
form (\ref{calv1}), is not a local operator.  This raises the question
of whether the tunneling operator can be relevant when $p>3$.

Here we show that there is indeed a local operator that will reduce
$\Delta_\sigma$, so that $2\Delta <2$, but that operator necessarily
also involves the charge sector.    Consider the charge $2e$ operator
introduced in (\ref{psi2er}), which is derived from $(\tilde
\psi_{ml}^\dagger)^2$.   In the bosonized representation this has the
form
\begin{equation}
\psi_{2e,R} \sim e^{i\pi N_\sigma} e^{i 4l \bar\phi_{\rho R}}
\partial_x\bar\phi_{\sigma R}.
\end{equation}
This motivates us to introduce a charge $2e$ tunneling term of the form
\begin{equation}
{\cal H}_{{\rm int}, i+1/2}' = \lambda \cos 4 l \bar\theta_{i+1/2,\rho}
\partial_x\bar\phi_{i,\sigma R}\bar\phi_{i+1,\sigma L}.
\end{equation}
When $t_4$ is relevant and flows to strong coupling, then
$\bar\theta_{i+1/2,\rho}$ is pinned at $\pi Q/4l$, where $Q$ is an
integer.    This leads to an effective interaction in the spin sector of
the form
\begin{equation}
{\cal H}_{{\rm int}, i+1/2}' = \lambda (-1)^Q
\partial_x\bar\phi_{i,\sigma R}\bar\phi_{i+1,\sigma L}.
\end{equation}
For fixed $Q$, this leads to a modification of the dimension
$\Delta_\sigma$.    Note that the sign of the interaction depends on
$Q$.   This reflects the fact that when $Q \rightarrow Q +1$, the spin
sector, $\cos( p\bar\theta_\sigma)$ and $\cos(p\bar\varphi_\sigma)$ are
interchanged in (\ref{pinsigma}).     Changing the sign of the
interaction takes $g_\sigma \rightarrow 1/g_\sigma$, which interchanges
the dimensions of $\cos( p\bar\theta_\sigma)$ and
$\cos(p\bar\varphi_\sigma)$.

\bibliography{refs}

\end{document}